\documentclass[
aps,%
12pt,%
final,%
notitlepage,%
oneside,%
onecolumn,%
nobibnotes,%
nofootinbib,%
superscriptaddress,%
noshowpacs,%
centertags,graphicx]%
{revtex4}

\usepackage[english]{babel}
\usepackage{graphicx,amsmath}

\begin{document}
\noindent {\it ASTRONOMY REPORTS, 2013, Vol 57}\bigskip \hrule  \vspace{10mm}

\title{Infrared Emission and the Destruction of Dust in HII
regions}

\author{\firstname{Ya.~N.}~\surname{Pavlyuchenkov
}}
\email{pavyar@inasan.ru}
\affiliation{Institute of Astronomy, Russian Academy of Sciences, Moscow, Russia
}%
\author{\firstname{M.~S.}~\surname{Kirsanova}}
\affiliation{%
Institute of Astronomy, Russian Academy of Sciences, Moscow, Russia
}%
\author{\firstname{D.~S.}~\surname{Wiebe}}
\affiliation{%
Institute of Astronomy, Russian Academy of Sciences, Moscow, Russia
}%

\begin{abstract}
{The generation of infrared (IR) radiation and the observed IR intensity distribution at wavelengths of 8, 24, and 100\,$\mu$m in the ionized hydrogen region around a young, massive star is investigated. The evolution of the HII region is treated using a self-consistent chemical-dynamical model in which three dust populations are included~-- large silicate grains, small graphite grains, and polycyclic, aromatic hydrocarbons (PAHs). A radiative transfer model taking into account stochastic heating of small grains and macromolecules is used to model the IR spectral energy distribution. The computational results are compared with \textit{Spitzer} and \textit{Herschel} observations of the RCW\,120 nebula. The contributions of collisions with gas particles and the radiation field of the star to stochastic heating of small grains are investigated. It is shown that a model with a homogeneous PAH content cannot reproduce the ring-like IR-intensity
distribution at 8\,$\mu$m. A model in which PAHs are destroyed in the ionized region provides a means to explain this intensity distribution. This model is in agreement with observations for realistic characteristic destruction times for the PAHs.}
\end{abstract}
\maketitle

\newpage

\section{Introduction}

One of the characteristic morphological features distinguished in infrared (IR) maps of the Galactic disk is so-called bubbles -- rings or fragments of rings in near-IR images (8-10\,$\mu$m), mainly obtained in the course of surveys on the \textit{Spitzer} and \textit{WISE} space telescopes. Some of these ring-like and pseudo-ring-like structures are associated with supernova remnants and planetary nebulae, although they are most often related to regions of ionized hydrogen in regions of formation of massive stars~\cite{c06,MWProject}.

IR bubbles typically have different spatial distributions in the near and middle-IR. As a rule, the inner region of a bubble in \textit{Spitzer} maps radiating at 24\,$\mu$m is surrounded by a ring of 8\,$\mu$m emission\cite{galbub1,galbub2}. The most natural explanation for this structure is projection of a three-dimensional shell onto the plane of the sky. The 8\,$\mu$m emission is usually taken to be associated with bands of polycyclic aromatic hydrocarbons (PAHs) excited by ultraviolet (UV) radiation~\cite{tielens}. It is logical to suppose that the PAH particles are concentrated in a shell, inside of which they are destroyed by the stellar radiation that drives the ionized hydrogen region (see, e.g.,~\cite{Verstraete}). The grains radiating at longer wavelengths, in particular at 24~$\mu$m, are larger in size, and so are able to survive at smaller distances from the star.

In addition to the destruction of grains, "mechanical"\, factors acting to move grains out of the central part of the HII region, such as radiation pressure and/or the action of the stellar wind, are also possible. An example is provided by the structure of the bubble N49 from the catalog~\cite{c06}. In this object, both the 8-$\mu$m and the 24-$\mu$m emission are distributed in rings. Everett and Churchwell~\cite{evch} have suggested that the inner ring outlines a cavity that has been swept out by the wind exciting the HII region. However, their estimates showed that this wind should sweep
out dust not only from the center, but from the entire HII region. To explain the presence of dust outside the wind-swept cavity, Everett and Churchwell~\cite{evch} considered the vaporization of small gas-dust clumps, or even protoplanetary disks, inside the HII region.

We investigate here possible explanations of the observed intensity distributions at wavelengths from 8 to 500\,$\mu$m in another well known object -- the IR bubble RCW~120~~\cite{rodgers:1960} -- taking into account only gas destruction processes. The ionized-hydrogen region RCW~120 is excited by a star whose spectral type is later than the star in N49; i.e., the star is less massive than the star in N~49, suggesting that the role of wind in the evolution of RCW~120 is not as great as in the case of N~49.

A fairly large number of recent papers have been dedicated to studies of the HII region RCW~120. This object has attracted attention for two reasons. First, it appears essentially spherically symmetric in \textit{Spitzer} and \textit{Herschel} images~\cite{deharveng:2005,zavagno:2010}. Second, star-forming regions are observed around RCW~120, which have apparently arisen during a collect-and-collapse process~\cite{zavagno:2007,deharveng:2009}. A massive protostellar object may be present in one of these "secondary"\ star-forming
regions~\cite{zavagno:2007}. Its relatively simple shape and the availability of varied observational data make RCW~120 a suitable target for our study.

\section{OBSERVATIONAL DATA}

\begin{figure}[t!]
\includegraphics[scale=0.45]{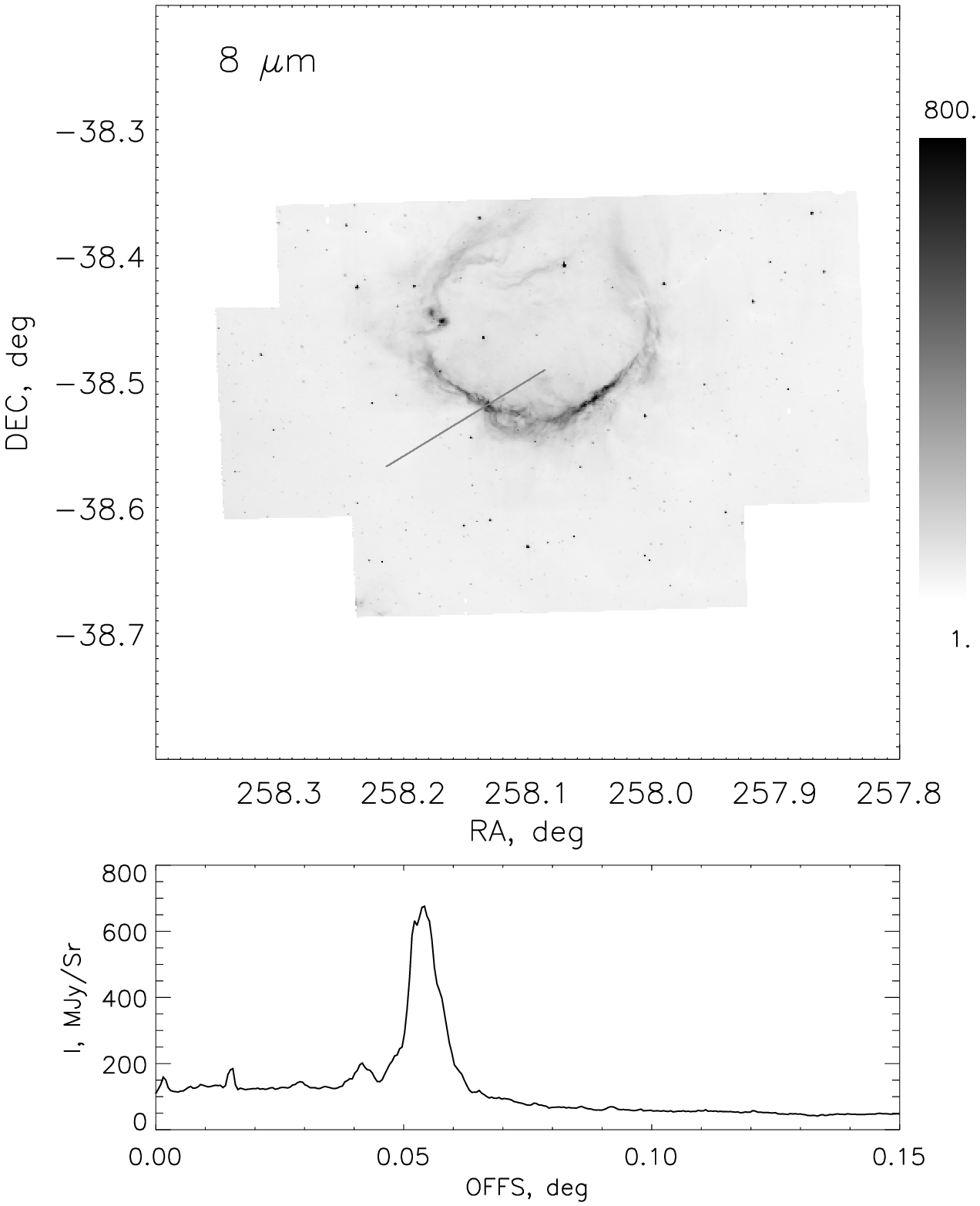}
\includegraphics[scale=0.45]{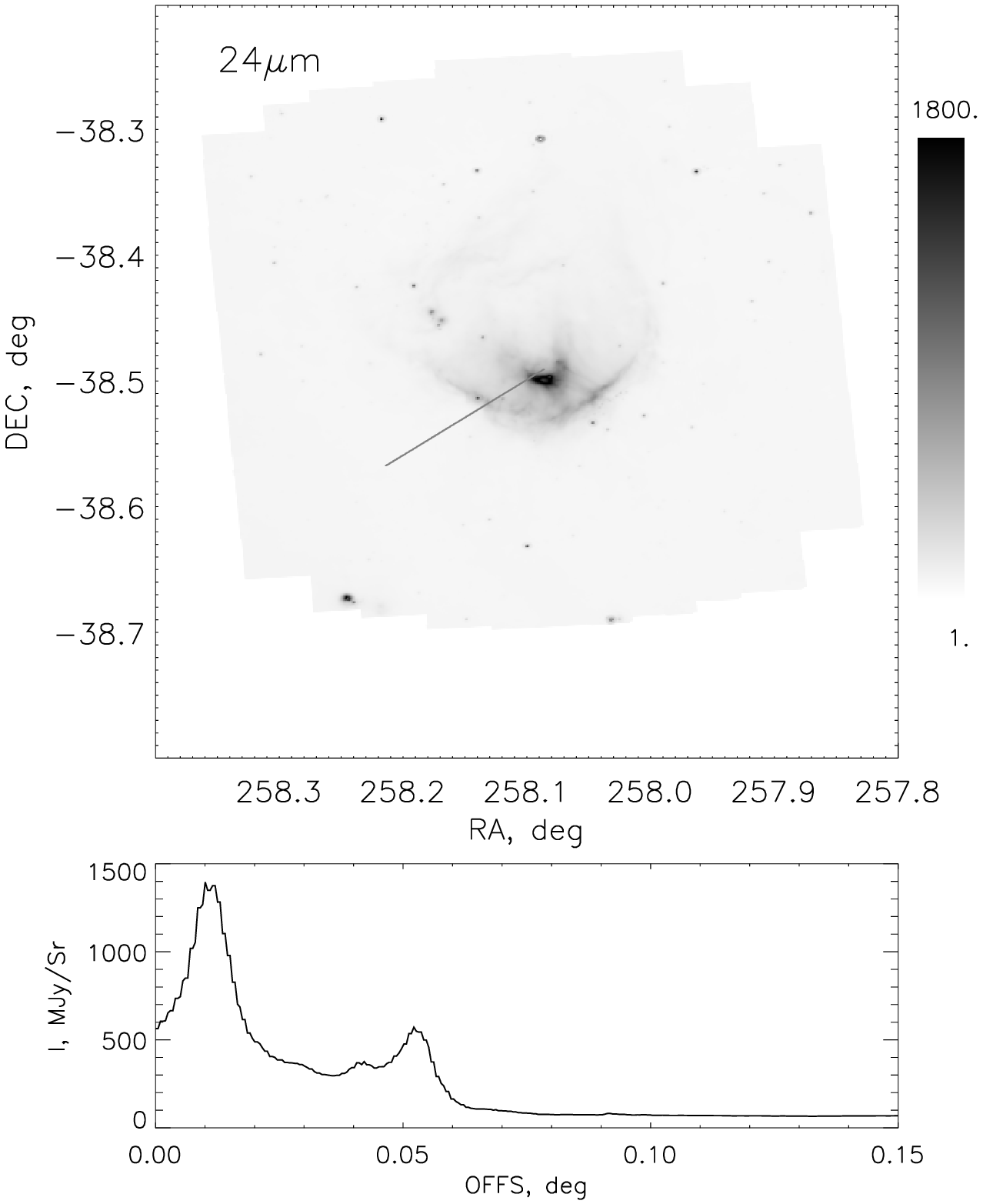}
\includegraphics[scale=0.45]{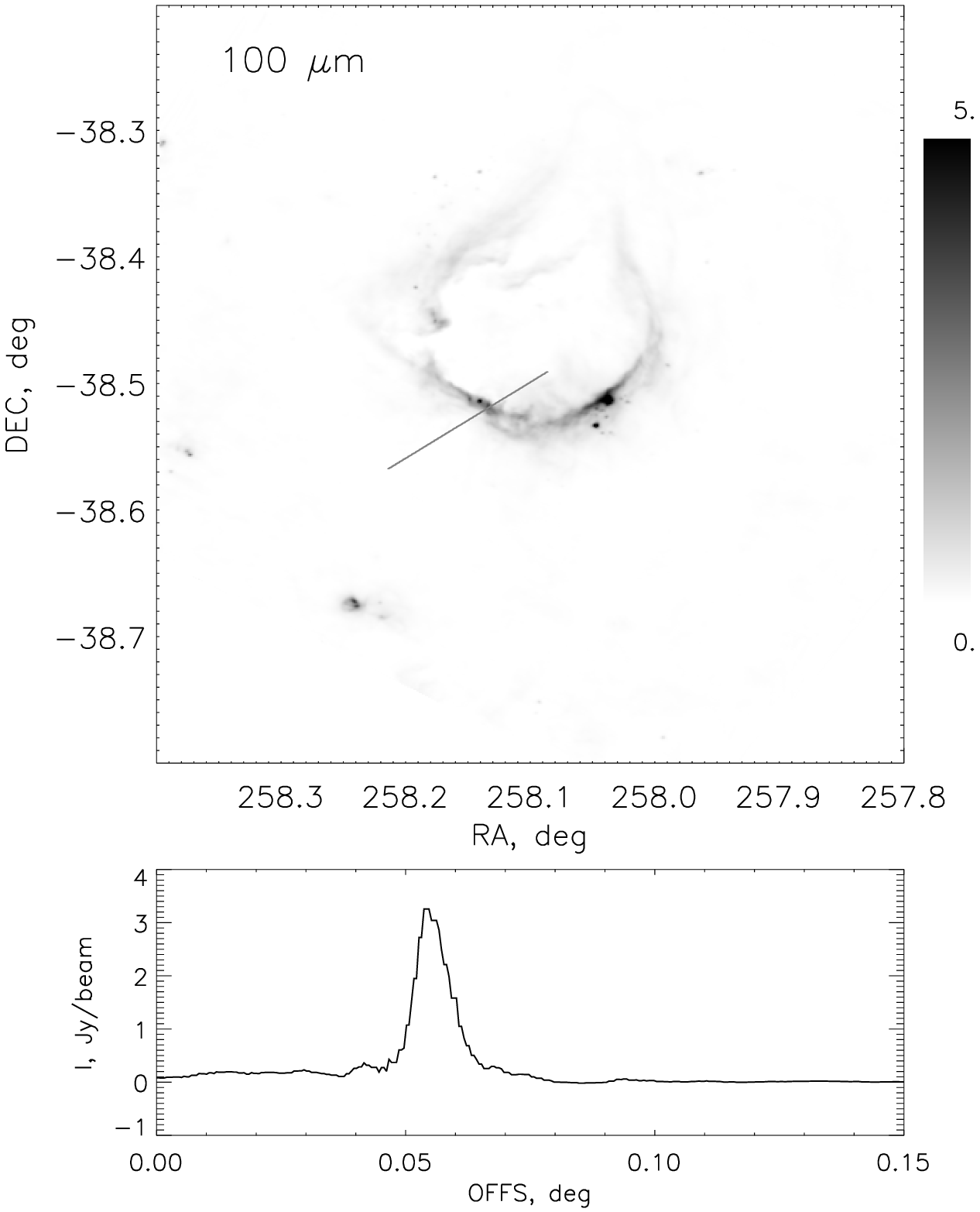}
\caption{Observed intensity maps of RCW~120 at $\lambda =$8\,$\mu$m (upper left), $\lambda =$24\,$\mu$m (upper right), and $\lambda =$100\,$\mu$m (lower). Intensity profiles along the lines shown in the maps are shown below each map. These data were taken from the archives of the \textit{Spitzer} and \textit{Herschel} space telescopes}
\label{obs}
\end{figure}

As was noted above, observations of RCW~120 were carried out on both the \textit{Spitzer} and \textit{Herschel} space telescopes. The \textit{Spitzer} observations were obtained during the GLIMPSE~\cite{Benjaminetal2003} and MIPSGAL~\cite{Careyetal2005} surveys, and are accessible from the Spitzer Heritage Archive\footnote{http://sha.ipac.caltech.edu}. The \textit{Herschel} observations of RCW~120 are described in~\cite{rcw120hersch} and are accessible from the Herschel Science Archive\footnote{http://herschel.esac.esa.int/Science\_Archive.shtml}. We used \textit{Spitzer} observations of RCW~120 at 8 and 24\,$\mu$m and \textit{Herschel} observations at 100\,$\mu$m. In all cases, the archival data were used without further processing. The observed maps of RCW~120 are presented in Fig.~\ref{obs}. Since the brightness of the ring varies appreciably with azimuth, we selected the brightness profile along a single direction for our analysis, indicated by the straight lines in Fig.~\ref{obs}.

Various estimates of the spectral type of the ionizing star CD--38$^\circ$11636 have been obtained: O8~\cite{georgelin:1970},
O9~\cite{crampton:1971}, and O6-8~V/III~\cite{martins:2010}. To take into account the scatter in these estimates, we obtained modeling
results for two effective temperatures for the ionizing
star: 30000 and 35000\,K~\cite{diaz_miller}. The gas density in
the cloud before the expansion of RCW~120 (the so called
initial density) is approximately estimated to
be 1400-3000~cm$^{-3}$ in~\cite{zavagno:2007}. This parameter was
varied in our models. The photometric distance to
RCW~120 is estimated to be 1.3$^{+0.4}_{-0.4}$\,kpc, and the
kinematic distance to be 1.8$^{+0.6}_{-0.7}$\,kpc~~\cite{russeil:2003}. We used the photometric distance when comparing our modeling with observations. The angular diameter of RCW~120 is 8$^\prime$~\cite{rodgers:1960}, which corresponds to a linear diameter of 3~pc at this distance.

\section{DYNAMICAL MODEL OF THE HII region}

We used the chemical-dynamical model described in detail by Kirsanova et al.~\cite{kirsanova:2009}, based on the Zeus2D software package~\cite{zeus}, when computing the physical structure of the HII region. This model can be used to trace the motion of ionization, dissociation, molecular-vaporization, and shock fronts during the expansion of the HII region into the surrounding molecular cloud. Although we are mainly concerned with the evolution of dust and modeling the radiation transfer, we must have a chemical model, since this enables us to correctly calculate cooling of the gas and treat the thermal balance in a self-consistent way at any moment in the evolution of the HII region.

The initial structure of the computational domain and the abundances of the various chemical components were the same as in~\cite{kirsanova:2009}. Further, we describe only those aspects of the modeling that differ from the original model. The key difference is related to the description of the grains. In the chemical-dynamical
model presented in~\cite{kirsanova:2009}, only a single dust component was included. In the extended version of the model we used, we assumed that the dust consists of three components: large silicate grains, very small grains (VSGs), and PAH particles. The parameters of these dust components are given in Table~\ref{tabledust}.

\begin{table}[htb]
\caption{Parameters of the dust components}
\bigskip
\begin{tabular}{c|c|c}
\hline
Component & Radius, cm & Density, g$\cdot$cm$^{-3}$ \\
\hline
Silicates       & $3\cdot 10^{-5}$ & 3.50 \\
Small grains    & $3\cdot 10^{-7}$ & 1.81 \\
PAHs            & $7\cdot 10^{-8}$ & 2.24 \\
\hline
\end{tabular}
\label{tabledust}
\end{table}

We assumed that icy mantles do not form on the VSGs and PAH particles. Such particles experience periodic heating to high temperatures (see below), which probably ''cleans''\ the surfaces of these particles of frozen molecules. However, the particles themselves can be destroyed by UV radiation from the star, ultimately forming atomic carbon (which subsequently participates in chemical processes). The initial contents of the grains $q_{\rm PAH}$ and $q_{\rm VSG}$, where $q_i$ is the fraction of PAHs or small graphite grains making up the overall atomic carbon abundance, are several percent. The exact values were chosen to provide a quantitative agreement between the computations and observations. It was assumed that the number of carbon atoms C in a single small graphite grain was $N_{\rm C}^{\rm VSG}\approx10^4$, and the number of carbon atoms in a PAH particle was $N_{\rm
C}^{\rm PAH}\approx150$. The relative initial
abundance of grains for dust type $i$ is then 
\begin{equation}
x_i(0)=(x_{\rm C}-x_{\rm C}^{\rm gas})q_i,
\label{eq:x}
\end{equation}
where $x_{\rm C}$ is the total relative abundance of carbon atoms and $x_{\rm C}^{\rm gas}$ is the initial relative abundance of C
atoms in the gas phase.

We modeled the destruction of small graphite grains using the expression
\begin{equation}
\dfrac{\partial x_{\rm VSG}}{\partial t} = -G \dfrac{x_{\rm  VSG}}{\tau_{\rm VSG}},
\label{eq:vsg}
\end{equation}
where $\tau_{\rm VSG}$ is the characteristic time for the destruction
of graphite grains in a standard UV radiation field and $G$ is the intensity of the stellar radiation in units of the Draine interstellar radiation field. We did not allow for the destruction of small grains when obtaining the results presented here. Equation~(\ref{eq:vsg}) is presented here only for completeness of the description of the model.

The equation used to model the evolution of the PAH particles contained a term describing the generation of PAH particles via the destruction of small graphite grains, together with a term describing the destruction of PAH particles:
\begin{equation}
\dfrac{\partial x_{\rm PAH}}{\partial t} = -G \dfrac{x_{\rm PAH}}{\tau_{\rm PAH}}
- \gamma \dfrac{\partial x_{\rm VSG}}{\partial t},
\label{eq:pah}
\end{equation}
where $\tau_{\rm PAH}$ is the characteristic time for the destruction
of PAH particles in a standard UV radiation field and $\gamma=\dfrac{N_{\rm C}^{\rm VSG}}{N_{\rm C}^{\rm PAH}}$. This definition of $\gamma$ means that, when small graphite grains are destroyed, all their mass is transformed into PAH particles. The destruction of PAH particles gives rise to carbon atoms, which enter
the gas phase. Accordingly, the term $G\dfrac{x_{\rm PAG}}{\tau_{\rm PAH}}N_{\rm C}^{\rm PAH}$ is added in the equation for $x_{\rm C}^{\rm gas}$ in the chemical model, to take into account the transformation of carbon atoms into the gas phase due to the destruction of PAH particles. All the types of grains are well mixed with the gas; i.e., participate in the advection together with gas in the hydrodynamic  computations.

We used the following results of the chemical-dynamical computation in the model for the radiative transfer and the construction of the theoretical intensity distributions: the number densities of neutral
hydrogen atoms H, molecular hydrogen H$_2$, H$^+$ ions, electrons, PAH particles, and small graphite grains, as well as the gas temperature and the mean spectral intensity of the radiation.

\section{RADIATIVE-TRANSFER MODEL}

The thermal state of the small grains can not be described
using an ordinary equilibrium temperature corresponding
to an average balance between absorption and emission processes. Their temperature is determined by single-photon heating, and can fluctuate over a wide range. In the general case, the thermal state of grain type $i$ was described using the probability-density distribution over the temperature, $P^{(i)}(T)$. Here, $P^{(i)}(T)dT$ is the fraction of grain
type $i$ having temperature in the range $(T,T + dT )$.
The NATA(LY) software package described in~\cite{pavlyuchenkov:2011,pavlyuchenkov:2012} was used to compute $P^{(i)}(T)$ and the radiation intensity distributions. The computational algorithm consists of the following steps.
\begin{enumerate}
\item Computation of the mean radiation intensity $J_\nu$ in each volume element of the cloud. We took the spatial distribution of $J_\nu$ from the chemical-dynamical model for the HII region
\item Computation of the probability density $P^{(i)}(T)$ in each volume element of the cloud. The computation of $P^{(i)}(T)$ was carried out by convolving the temperature history of an individual grain located in
gas with specified physical parameters in a specified radiation field $J_\nu$. The temperature history of an individual grain, in turn, was calculated using the Monte Carlo method to model the discrete heating of the grain due to the absorption of UV photons and collisions with gas particles, and cooling of the grain via its own IR emission.\
\item Computation of the spectrum of the emergent radiation via integration of the transfer equation along a chosen direction. Emission, absorption, and isotropic scattering were taken into account when modeling the spectra. The emission coefficient $j_\nu$ in each cell of the cloud was calculated using the derived distributions $P^{(i)}(T)$.
\end{enumerate}
The absorption and scattering efficiencies for silicate and graphite grains required for the modeling were calculated using Mie theory applied to spherical grains of the corresponding sizes. The absorption efficiencies for the PAHs were calculated using formulas from~\cite{draine:2007}. The parameters describing the heat capacity of the grains were similar to those used in~\cite{pavlyuchenkov:2012}.

We considered two mechanisms for stochastic heating of the grains: absorption of UV photons and collisions with gas particles. The method used to calculate the probability density $P^{(i)}(T)$ for heating
by UV radiation is described in detail in an appendix to~\cite{pavlyuchenkov:2012}. The algorithm used to calculate $P^{(i)}(T)$ for the case of heating by gas particles differs only in that it generates a sequence of collisions $\lbrace E_1, . . . ,E_M \rbrace$, where $E_j$ is the kinetic energy of a gas particle (atom, ion, or electron) colliding with a grain, rather than a sequence of photon absorptions $\lbrace h\nu_1, . . . , h\nu_M\rbrace $ by the grain. The sequence of collision energies for a Maxwellian velocity distribution for the gas particles
is generated using the density distribution
\begin{equation}
f(E)=\frac{4}{3\sqrt\pi} \left( \frac{E}{kT}\right)^{\frac{3}{2}} \exp(-E/kT) 
\end{equation}
The corresponding sequence of collision times $\lbrace t_1, . . . , t_M\rbrace $ satisfies a Poisson flow of events,
\begin{equation}
f(t)=\lambda \exp(-\lambda t),
\end{equation}
where $f(t)$ is the density function of the distribution of times between successive collisions and $λ = M/t_{\rm 0}$ is the mean number of collisions per unit time. The total time $t_{\rm 0}$ is found from the relation
\begin{equation}
\frac{M}{t_{\rm 0}}=\pi a^2 \sum_j n_j \overline{V}_j,
\end{equation}
where $n_j$ is the number density and $\overline{V}_j = \sqrt{\dfrac{8kT}{\pi m_j}}$ is the mean thermal velocity of particles of type $j$.

We assumed that the kinetic energy of the particles colliding with the grains is converted fully into heating of the grains. The heating of the dust by the gas is obviously overestimated in this case. Note that, in this approximation, the largest contribution to heating in the hydrogen-ionization region is made by free electrons.

\section{RESULTS FOR MODELS WITHOUT DESTRUCTION OF DUST}

Before turning to processes that can destroy dust particles, let us consider the main characteristics of the evolution of the observed manifestations of an HII region containing several dust populations. We considered three models with different initial gas densities
in the parent molecular cloud ($n_{\rm 0}$) and effective temperatures for the ionizing star ($T_{\rm eff}$). These models assumed that neither large dust particles or PAH or VSG particles were destroyed, and that they only moved together with the gas. Information about these models without the destruction of dust is presented in Table~\ref{nodestruction_models}. Our basic model had $T_{\rm eff} = 35 \times 10^3$~K and $n_{\rm 0} = 3 \times 10^3$~cm$^{−3}$ (model 3e3-35).

\begin{table}[h!]
\caption{Models without destruction of dust}
\bigskip
\begin{tabular}{c|c|c|c|c|c}
\hline
Model & $T_{\rm eff}$, K & $n_{\rm 0}$, cm$^{-3}$ &\multicolumn{3}{c}{Mass fraction of dust}\\ \cline{4-6}
       &                  &                        & Silicates & VSG & PAH\\
\hline
3e3-35 & 35$\cdot 10^3$ & 3$\cdot 10^3$ & 94.6\% & 3.6\% & 1.8\%\\
3e3-30 & 30$\cdot 10^3$ & 3$\cdot 10^3$ & 90.7\% & 5.6\% & 3.7\%\\
1e4-35 & 35$\cdot 10^3$ & 1$\cdot 10^4$ & 94.6\% & 3.6\% & 1.8\%\\
\hline
\end{tabular}
\label{nodestruction_models}
\end{table}

\subsection{Physical Properties of the Gas near an Expanding HII region}

\begin{figure}[t!]
\includegraphics{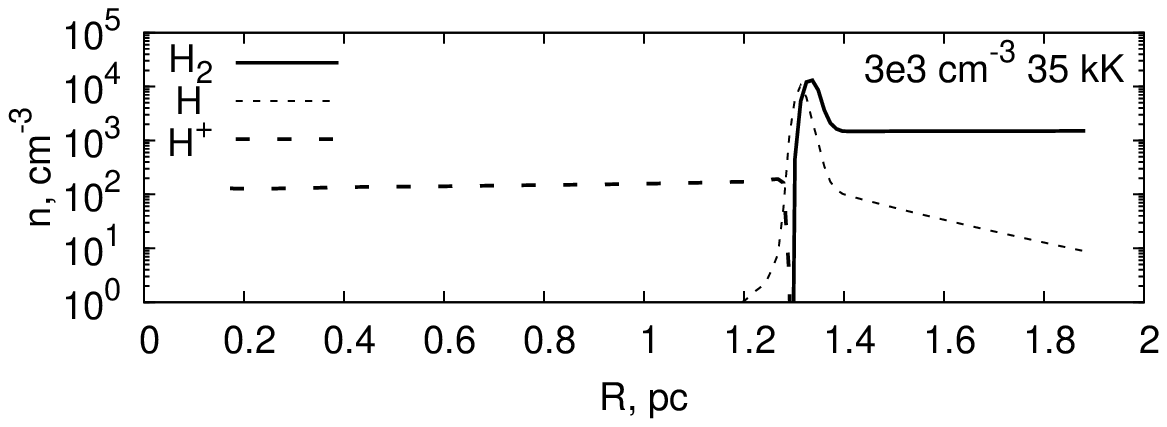}\\
\includegraphics{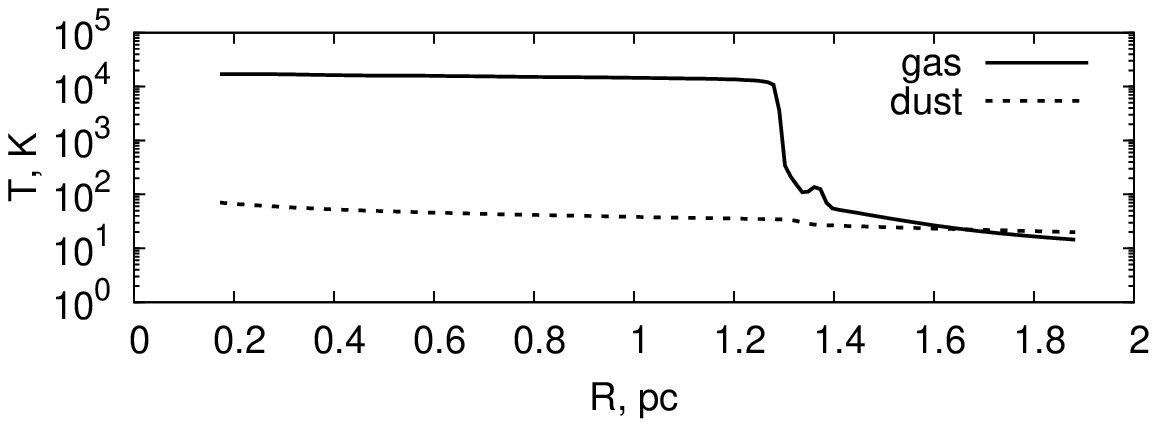}
\caption
{Distributions of the number density (upper) and temperature (lower) for a model HII region with $n_{\rm 0} = 3000$~cm$^{−3}$ and
$T = 35 000$~K.}
\label{model}
\end{figure}

Figure~\ref{model} shows the physical properties of the gas and dust for the basic model (3e3-35) inside the HII region, in the surrounding molecular cloud, and in a relatively thin, dense transition layer between them. This figure corresponds to a time $t = 170 000$~yrs after the onset of the expansion of the HII region. At this time, the diameter of the model HII region was close to the real diameter for the adopted distance. The electron density is also in agreement with the observed value, 86\,cm$^{-3}$~\cite{zavagno:2007}. The upper plot in Fig.~\ref{model} shows the density distributions of ionized, atomic, and molecular hydrogen. Both atomic and molecular hydrogen are located in the dense layer bounding the shock front on the side of the molecular cloud and the ionization front on the side of the HII region. The density of atomic hydrogen falls off with distance from the HII region in the molecular cloud. The width of the dense layer is 0.1~pc, which comprises about 10\% of the size of the HII region.

The gas temperature exceeds the dust temperature inside the HII region, in the dense transition layer, and, partially, in the molecular cloud. The difference is maximum inside the HII region: the gas temperature is of order 10$^4$~K, while the dust temperature does not exceed 100~K. There is a drop in the gas temperature by two orders of magnitude in the dense layer.

\begin{figure}[t!]
\includegraphics{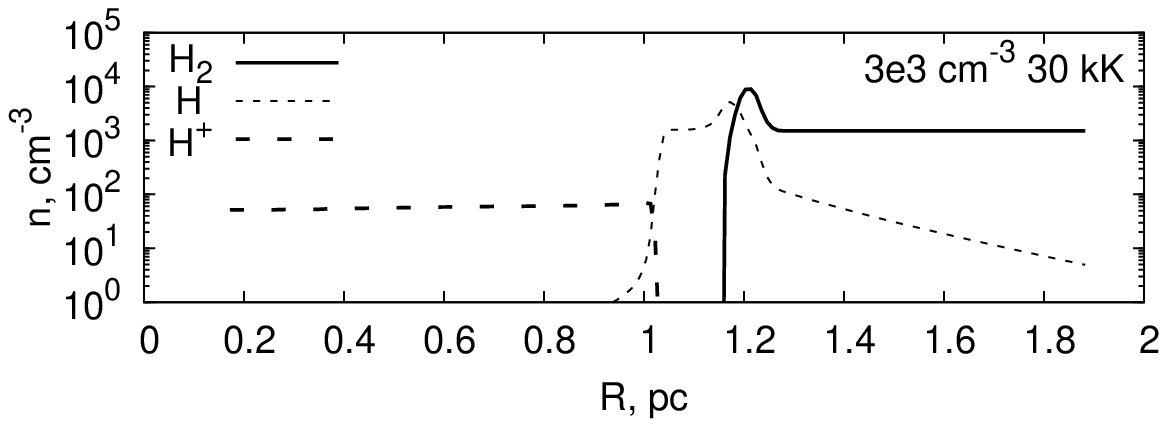}\\
\includegraphics{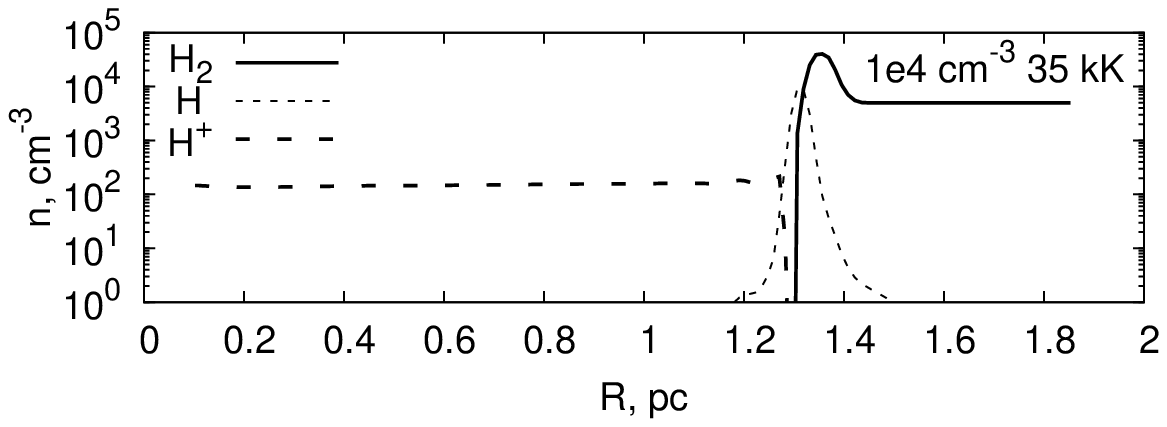}
\caption{Density distributions in model HII region with various initial densities and stellar temperatures (indicated to the upper right in the plots)}
\label{model_alt}
\end{figure}

Figure~\ref{model_alt} shows how the physical structure of the region surrounding the HII region varies with the effective temperature of the ionizing star (upper panel, model 3e3-30) and the initial gas density (lower panel, model 1e4-35). The results for models 3e3-30 and 1e4-35 are shown for times of 280000~yrs and 320000~yrs after the onset of the expansion of the HII region. In both cases, these are the times when the size of the model HII region corresponds to the observed size.

This figure shows that the layer of atomic hydrogen between the HII region and the molecular cloud is more extended in the HII region around the latertype star, for the same size HII region. In particular, this layer is wider in model 3e3-30 than in model 3e3-35. The width of the atomic-hydrogen layer is determined by the ionization rate, on the one hand, and the molecular-hydrogen dissociation rate, on the other. The flux of UV radiation ionizing the hydrogen is a factor of 11 higher for a star with $T_{\rm eff} = 35\times10^3$~K
than for a star with $T_{\rm eff} = 30\times10^3$~K, but the flux of softer UV radiation capable of dissociating the molecular hydrogen is only a factor of two higher.

For an initial gas density 10$^4$ cm$^{−3}$ and the same effective temperature as in the former model ($T_{\rm eff} = 35\times10^3$~K), the width of the dense layer between the HII region and the unperturbed molecular gas again becomes of order 0.1 pc. In contrast to the basic model, the bulk of the mass of the dense layer is in
the form of molecular hydrogen. The density of atomic hydrogen falls sharply inside the dense layer. The high initial gas density in model 1e4-35 prevents the UV radiation from penetrating inside the dense layer and dissociating the molecular hydrogen there.

\subsection{Contribution of UV Radiation and Hot Gas to Stochastic Heating of the Dust}

The gas temperature in the region of ionization is about 14000~K. This is appreciably higher than the dust temperature in the model HII region calculated in the approximation of radiative equilibrium (about 50~K). The corresponding internal energy of the small grains,
$E \approx 3NkT \approx 1$~eV (where $N \approx 100$ is the characteristic number of atoms in a grain), is comparable to the kinetic energy of a gas particle ($\approx1$~eV). This means that the exchange of energy between gas particles and dust grains should occur in a stochastic regime.

To illustrate the contribution of the gas to stochastic heating of the dust, and compare this contribution to that from heating by the UV radiation field, we present results of our computations of the thermal
structure and spectra for models in which only one or the other of these mechanisms were included. We chose model 3e3-35 as the model for the HII region.

\begin{figure}[t!]
\includegraphics[scale=0.4]{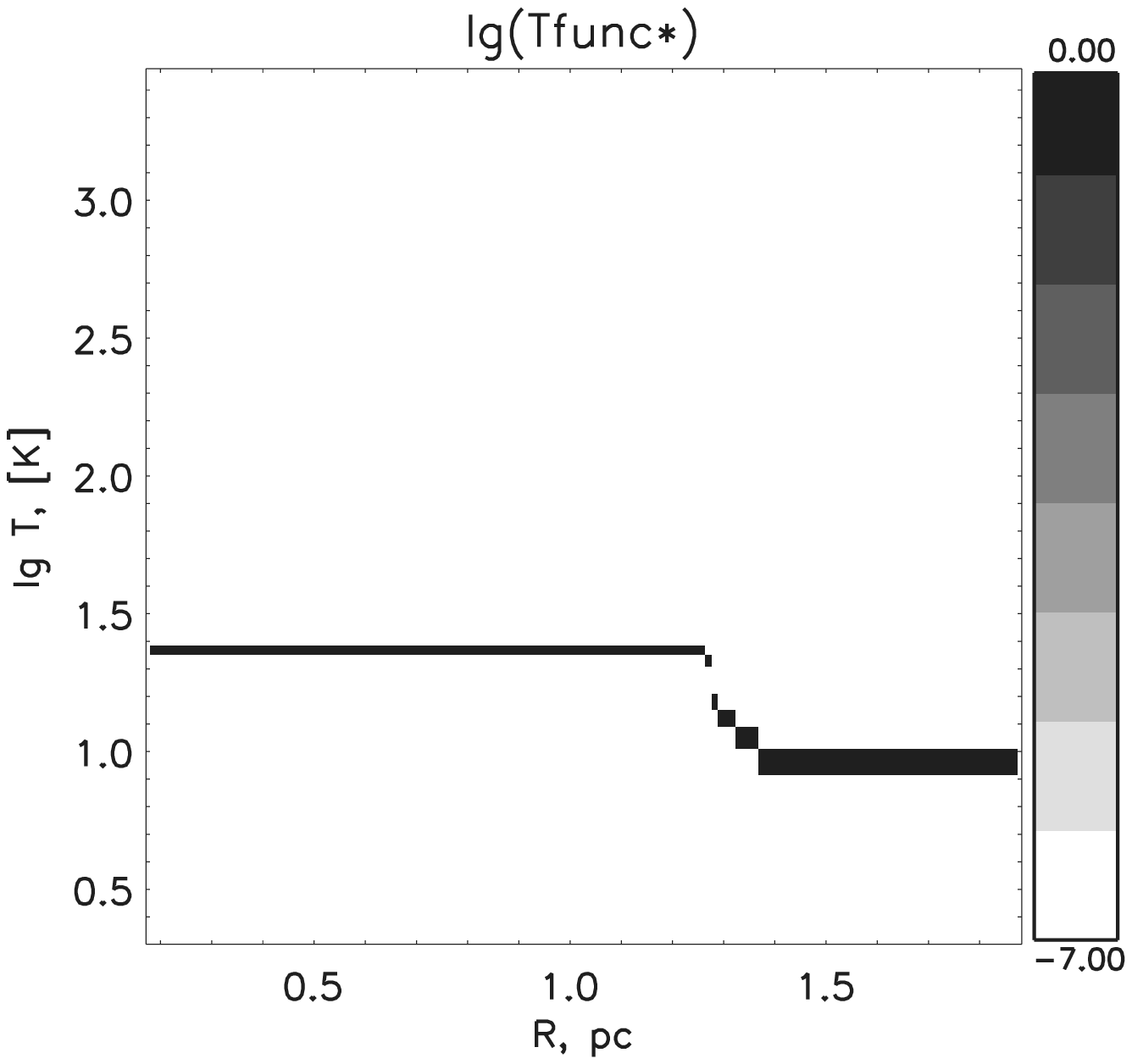}
\includegraphics[scale=0.4]{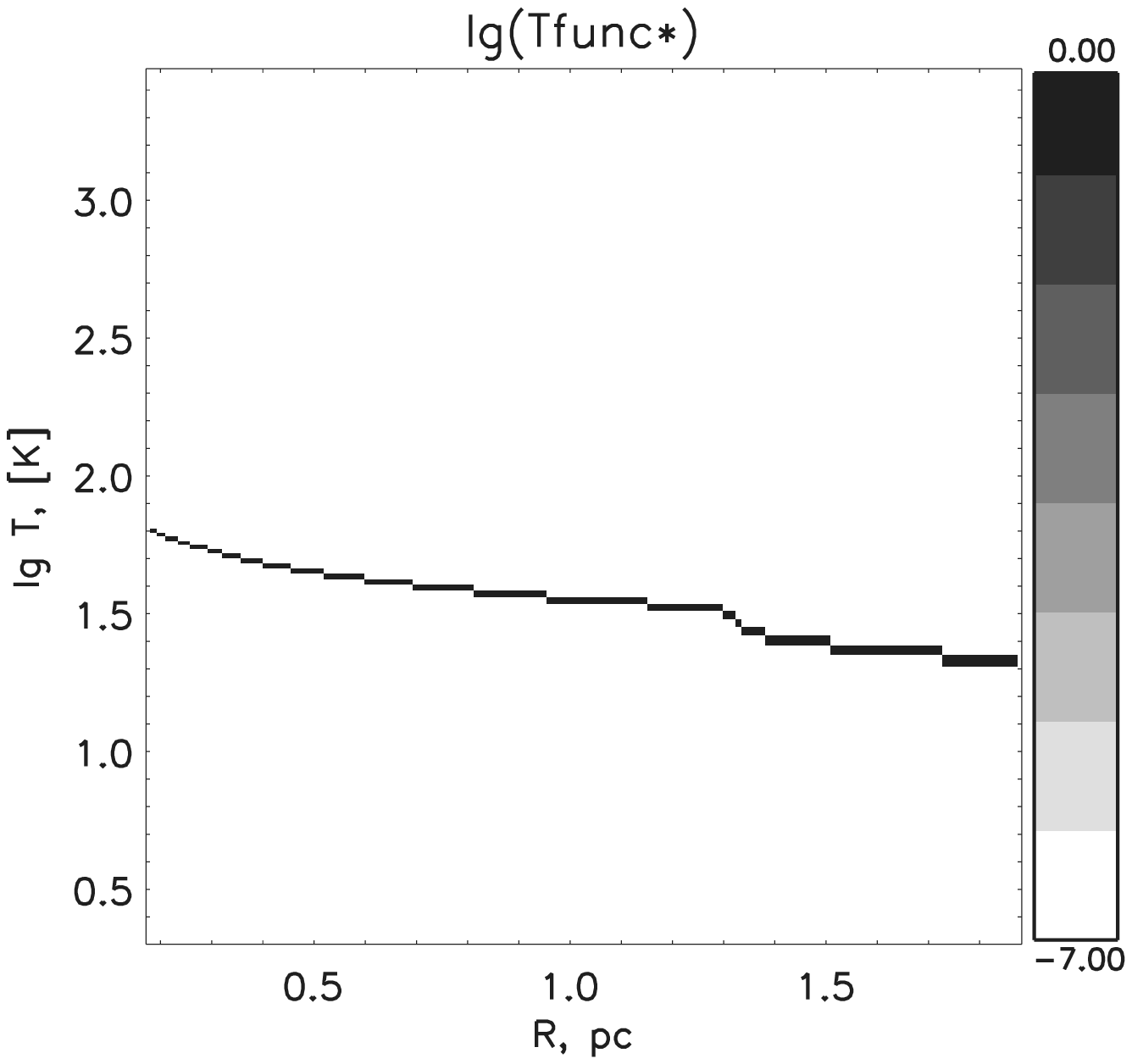}\\
\includegraphics[scale=0.4]{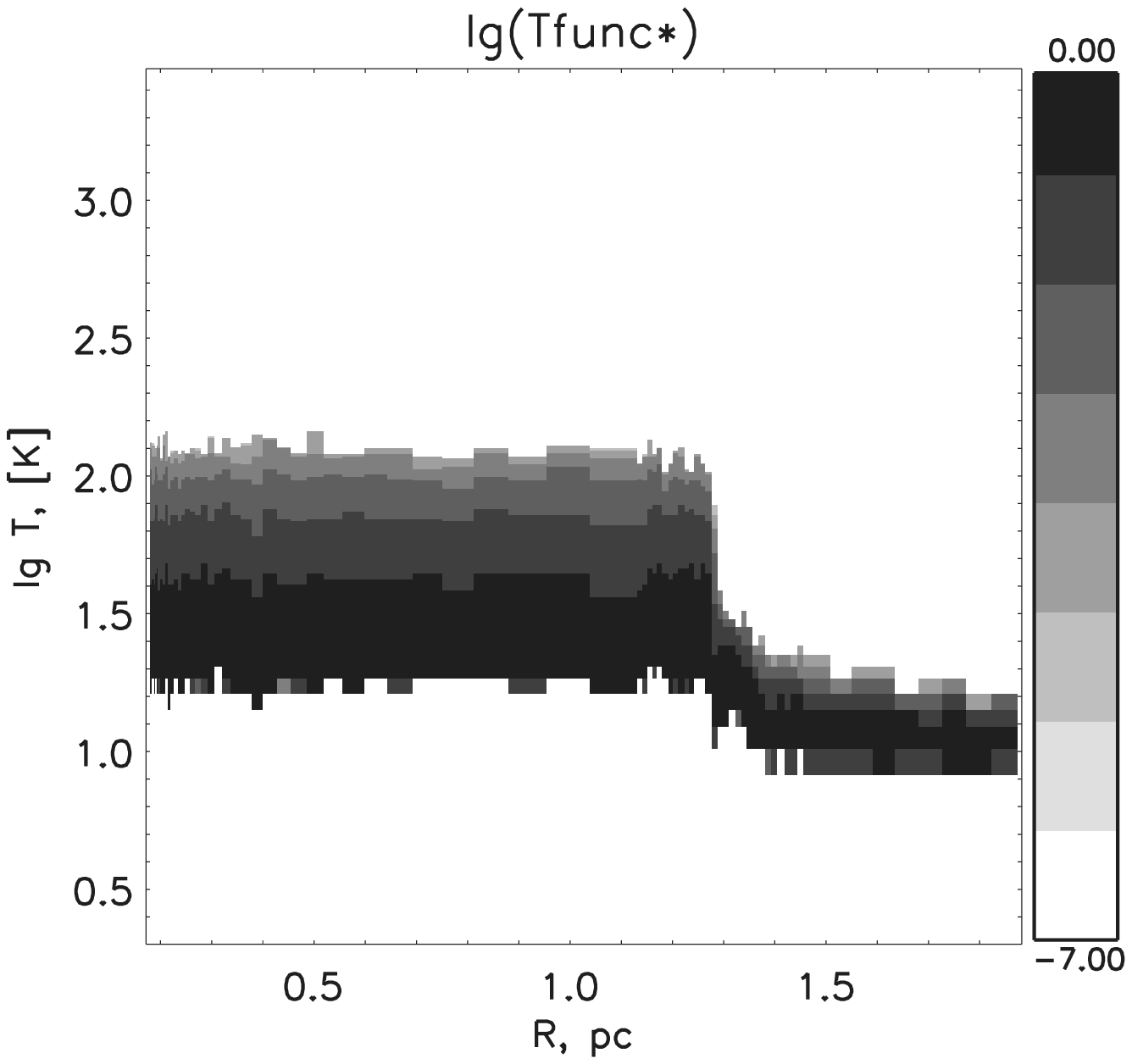}
\includegraphics[scale=0.4]{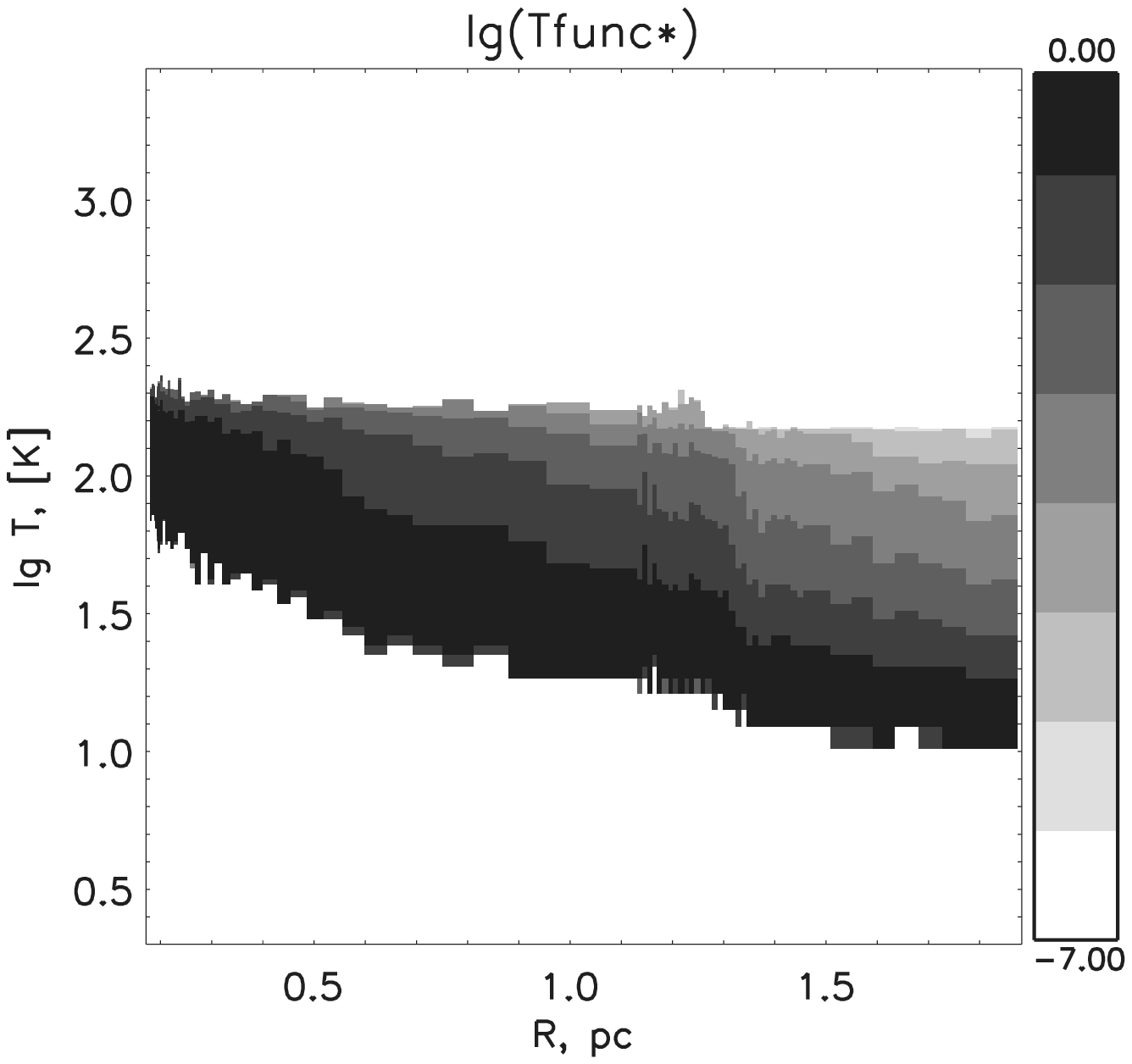}\\
\includegraphics[scale=0.4]{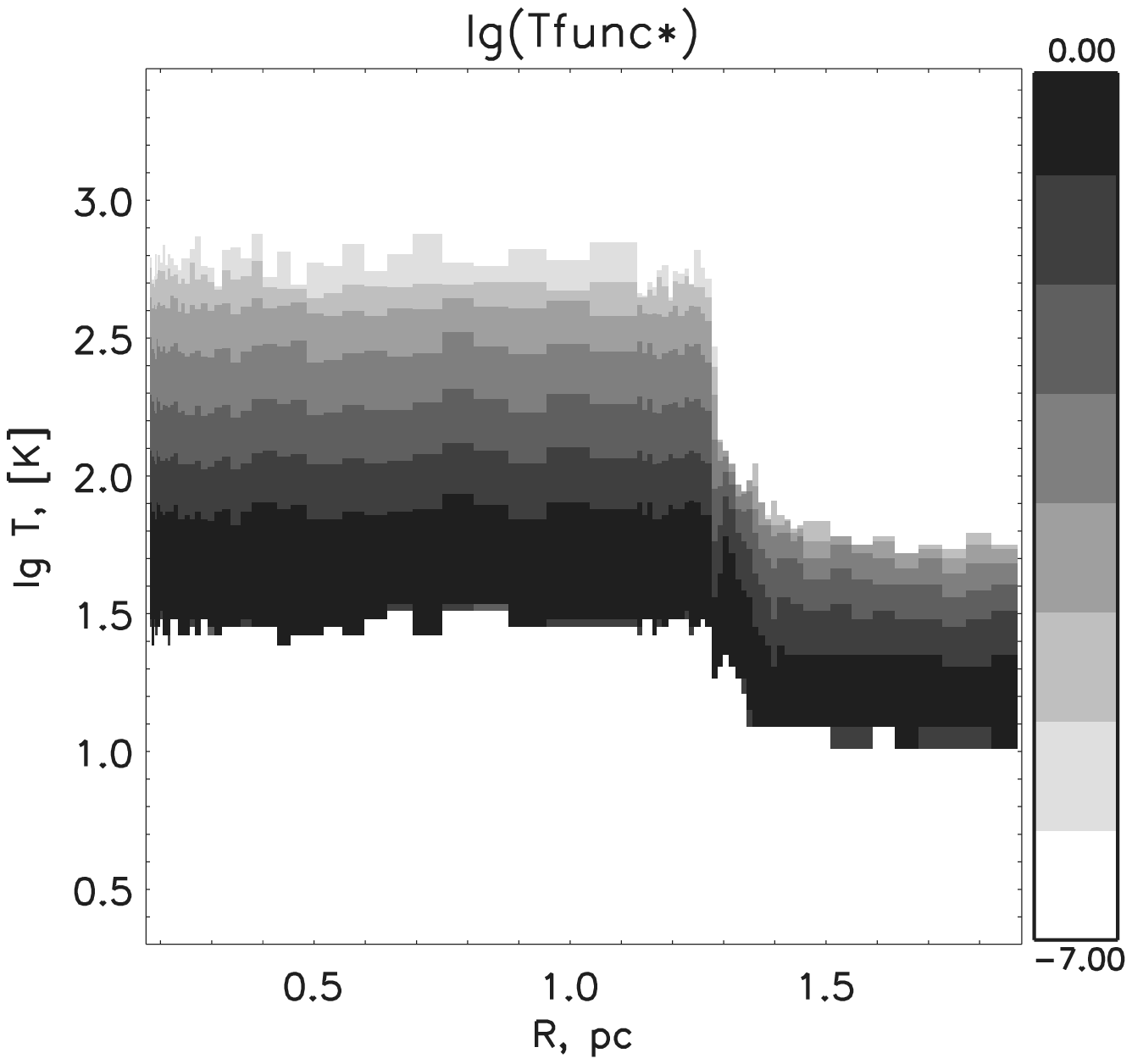}
\includegraphics[scale=0.4]{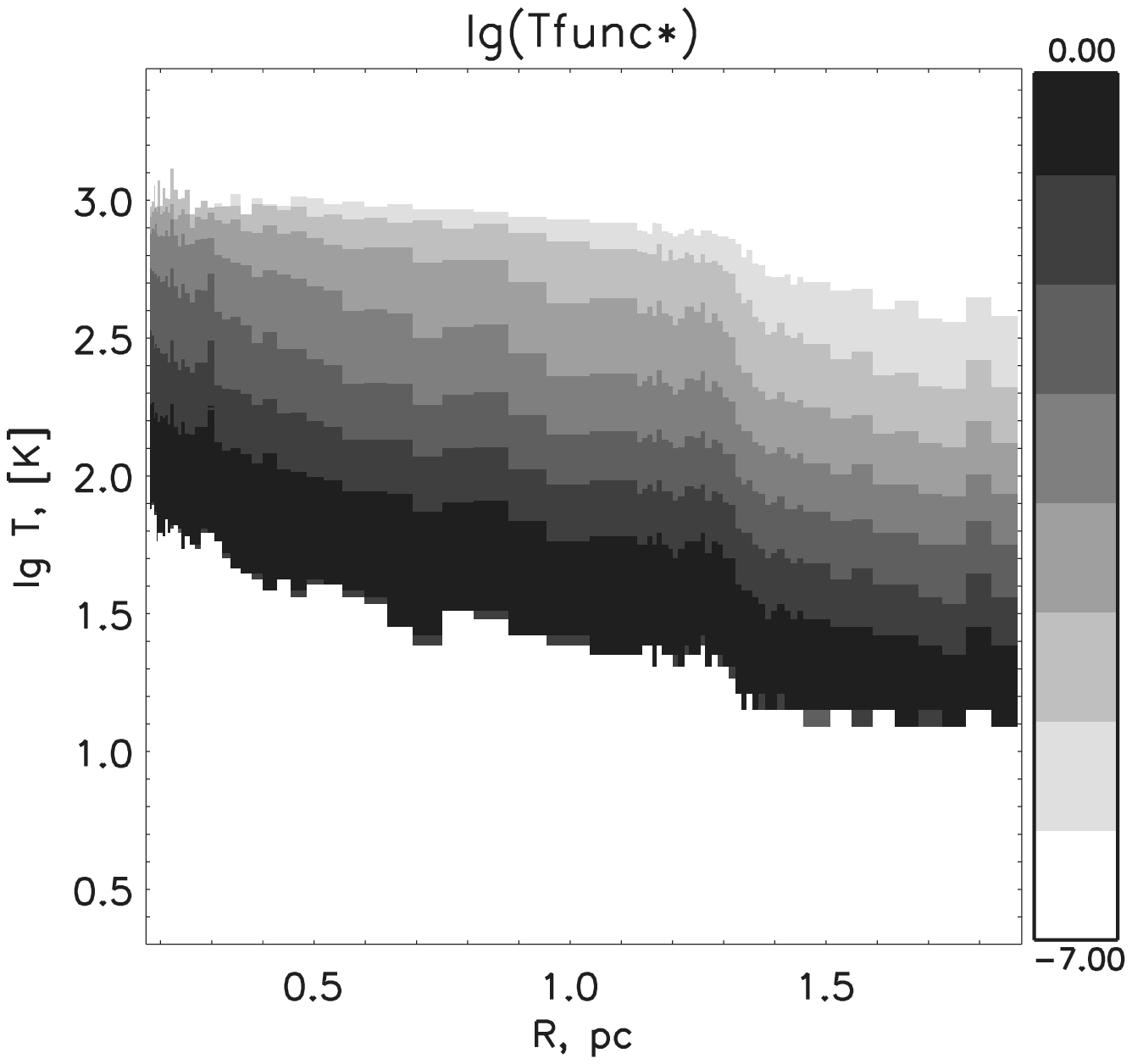}\\
\caption{Spatial distribution of the normalized probability density for various dust-temperature regimes: when only collisions with ions, atoms, and electrons (left column) and when only the UV radiation field (right column) are taken into account when modeling the stochastic heating of the dust. The upper, middle, and lower pairs of panels show results for large silicate grains, small graphite grains, and PAH particles, respectively.}
\label{funTemp}
\end{figure}

\begin{figure}[t!]
\includegraphics[scale=0.5]{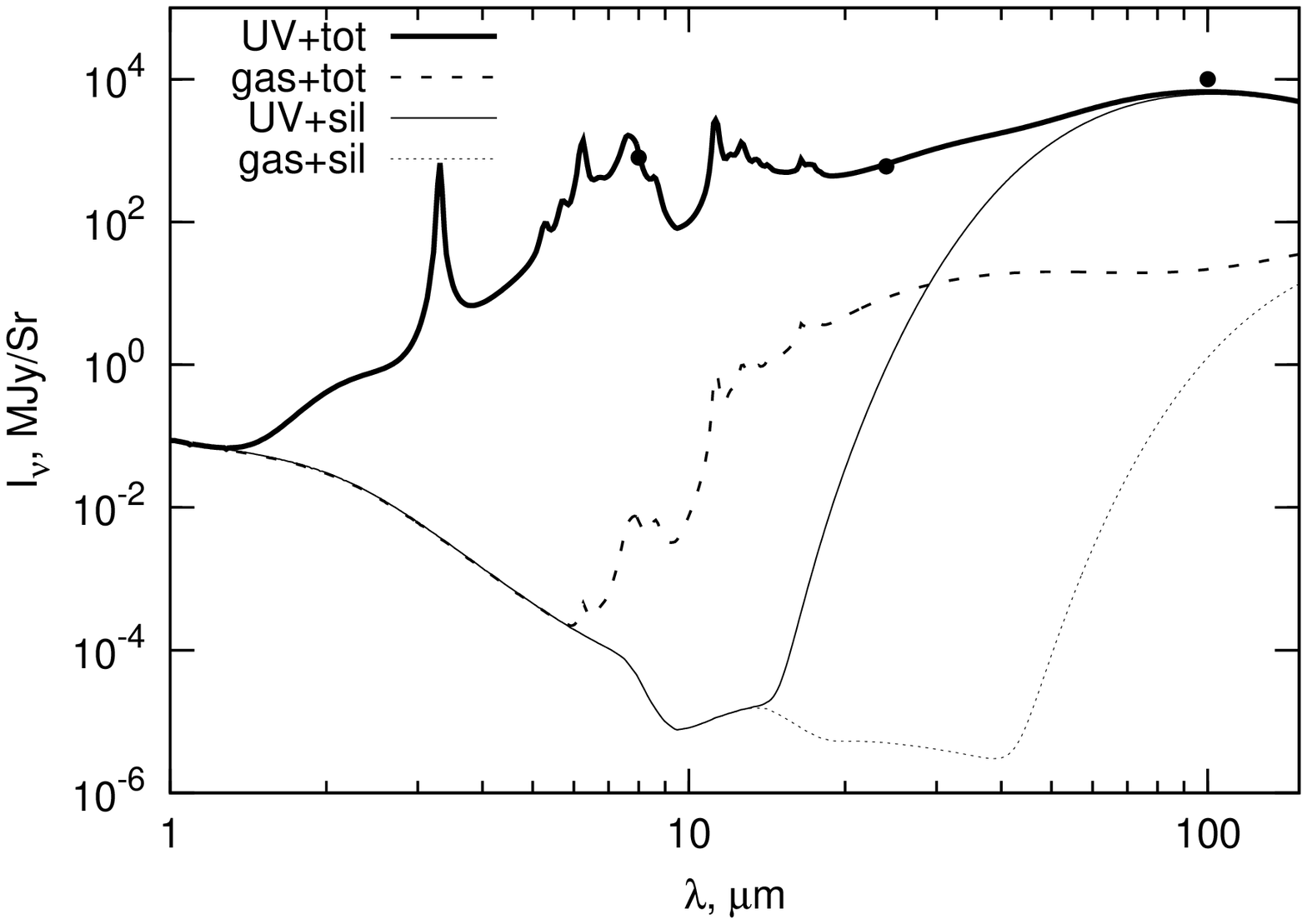}
\caption{Spectra along the direction toward the ring for various dust-heating models, taking into account all the dust components, with dust heating only by the UV radiation field (UV + total; bold solid curve); only large silicate grains, with dust heating only by the UV radiation field (UV~+~sil; thin solid curve); all the dust components, with heating only by collisions with atoms, ions, and electrons (gas~+~total; dashed curve); and only large silicate grains, with heating only by collisions with atoms, ions, and electrons (gas~+~sil; dotted curve). The points show the observed intensities for RCW~120.}
\label{ColTemp}
\end{figure}

Dependences of the normalized probability distribution
$\tilde{P}(T)=P(T)/\max\{P(T)\}$ on distance from the star for the grains considered are shown in Fig.~\ref{funTemp}. Both heating mechanisms lead to fluctuations in the temperature of the small grains and PAH particles, as is reflected by the spreading of the $\tilde{P}(T)$ distribution in the vertical direction. At the same time, the temperature of the large grains does not fluctuate, and the $\tilde{P}(T)$ dependence degenerates into a radial dependence for the temperature. Overall, the model with UV heating of the dust has a higher temperature than the model with collisional heating. Figure~\ref{ColTemp} shows corresponding spectra in the direction toward the ring of emission at 8 $\mu$m. The intensity in the models with UV heating (solid curves) exceeds the intensity in the corresponding models with collisional heating (dashed and dotted curves) by more than an order of magnitude over the entire spectrum. This led us to not consider further heating of the dust via energy exchange with the gas. Note also that the model with only large grains (thin solid curve) is not able to reproduce the observed spectrum in the near IR, even when UV heating is included.

\subsection{Theoretical Spectra in the IR}

\begin{figure}[t!]
\includegraphics[scale=0.5]{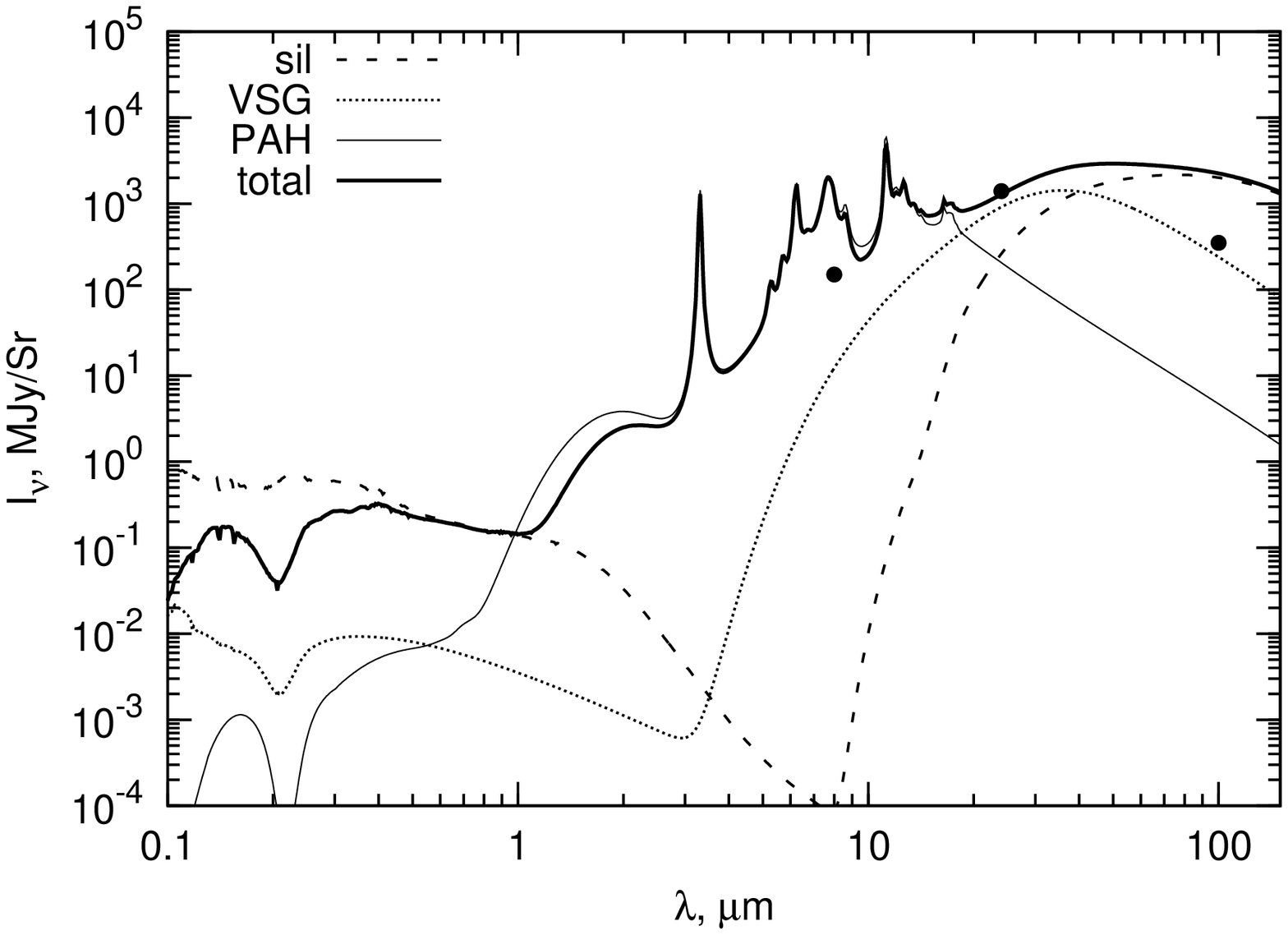}\\
\includegraphics[scale=0.5]{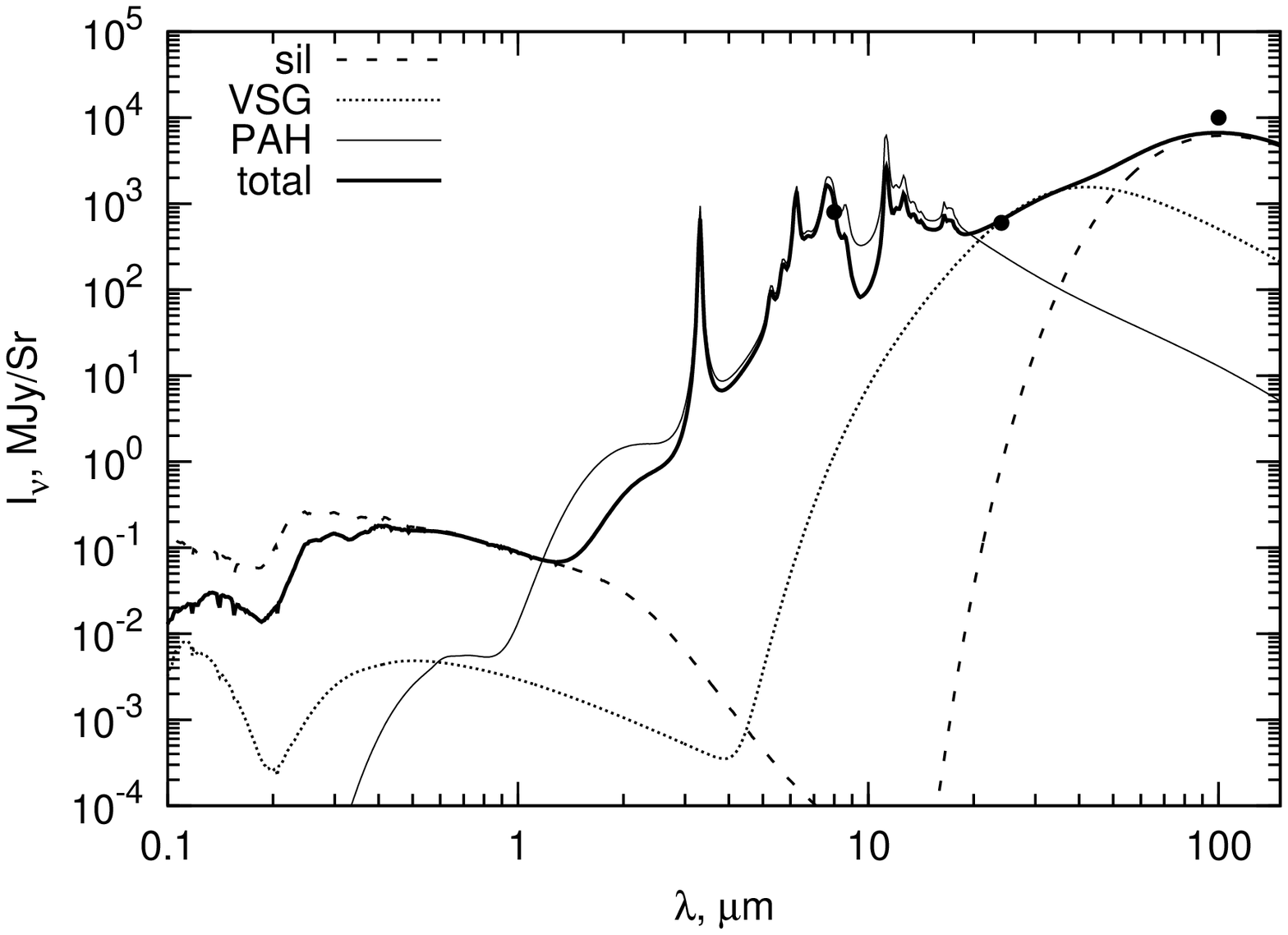}
\caption{Spectra for the model with UV heating of the dust, in the direction toward the center of the cloud (upper) and in the direction toward the ring (lower). The spectra formed by the individual dust components (sil~-- large silicate grains, VSG~-- very small grains, PAH~-- polycyclic aromatic hydrocarbons) and by all the components together (total) are shown by the different types of curves. The points show the observed intensities for RCW~120.}
\label{SED}
\end{figure}

Figure~\ref{SED} shows the contributions of individual dust components to the spectrum for the model with UV heating. The main contribution at wavelengths $\lambda > 40 \mu$m is made by thermal radiation by large grains (dashed curves), while radiation from small graphite grains dominates the intensity at $20 \mu{\rm m} < \lambda < 40 \mu{\rm m}$, and stochastically heated PAH particles dominate the spectrum at $1 \mu{\rm m} < \lambda < 20 \mu{\rm m}$. The spectrum at $\lambda < 1 \mu{\rm m}$ is determined by scattering of interstellar radiation on large grains. The solid bold curves show the combined spectrum for all the components. Note that these spectra lie below the spectrum formed by large grains at $\lambda < 1 \mu{\rm m}$. This is due to the fact that including all the dust components leads to higher absorption at these wavelengths; i.e., radiation scattered by large grains is efficiently absorbed by the smaller grains, especially the PAH particles.

\begin{figure}[t!]
\includegraphics[scale=0.42]{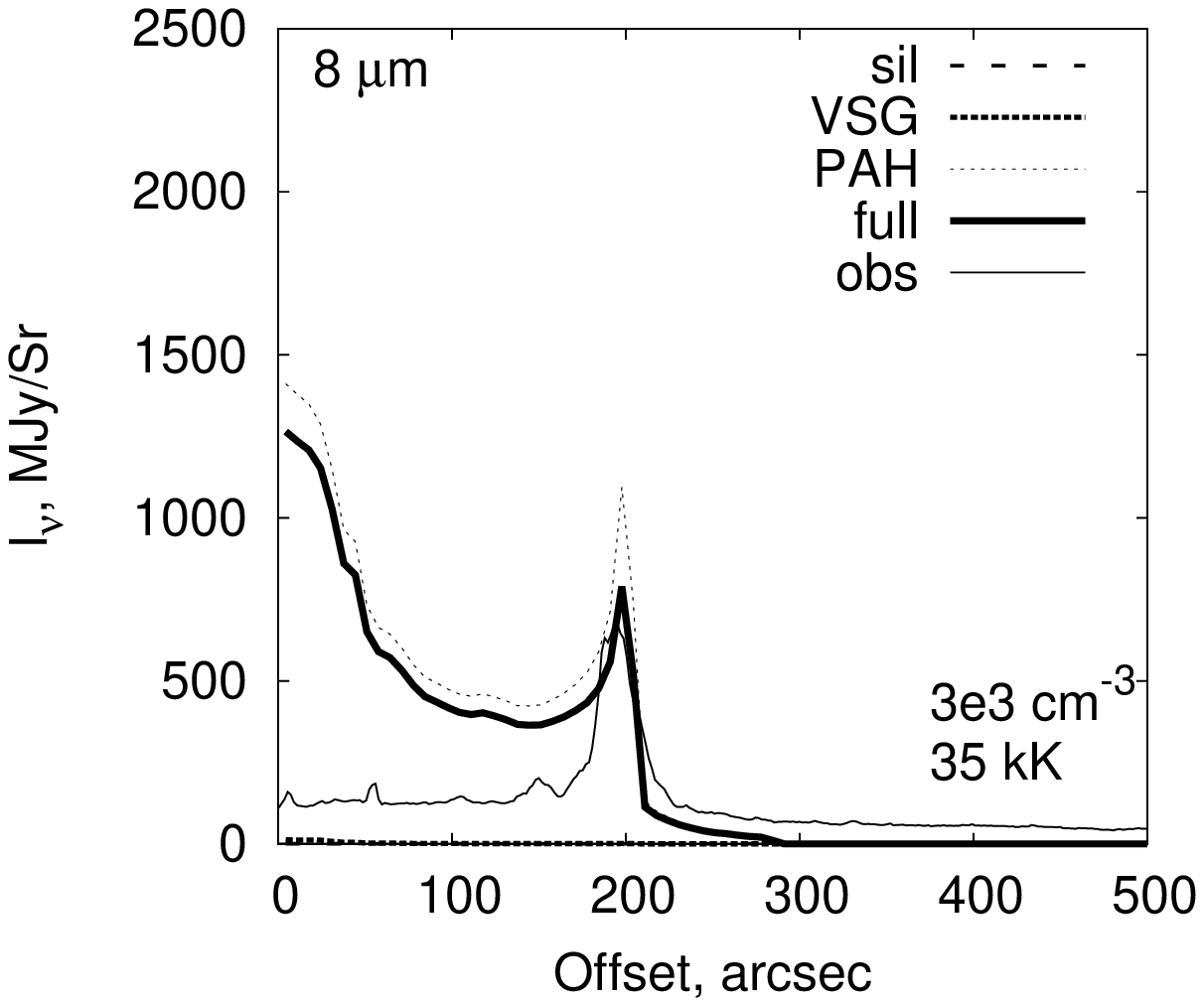}
\includegraphics[scale=0.42]{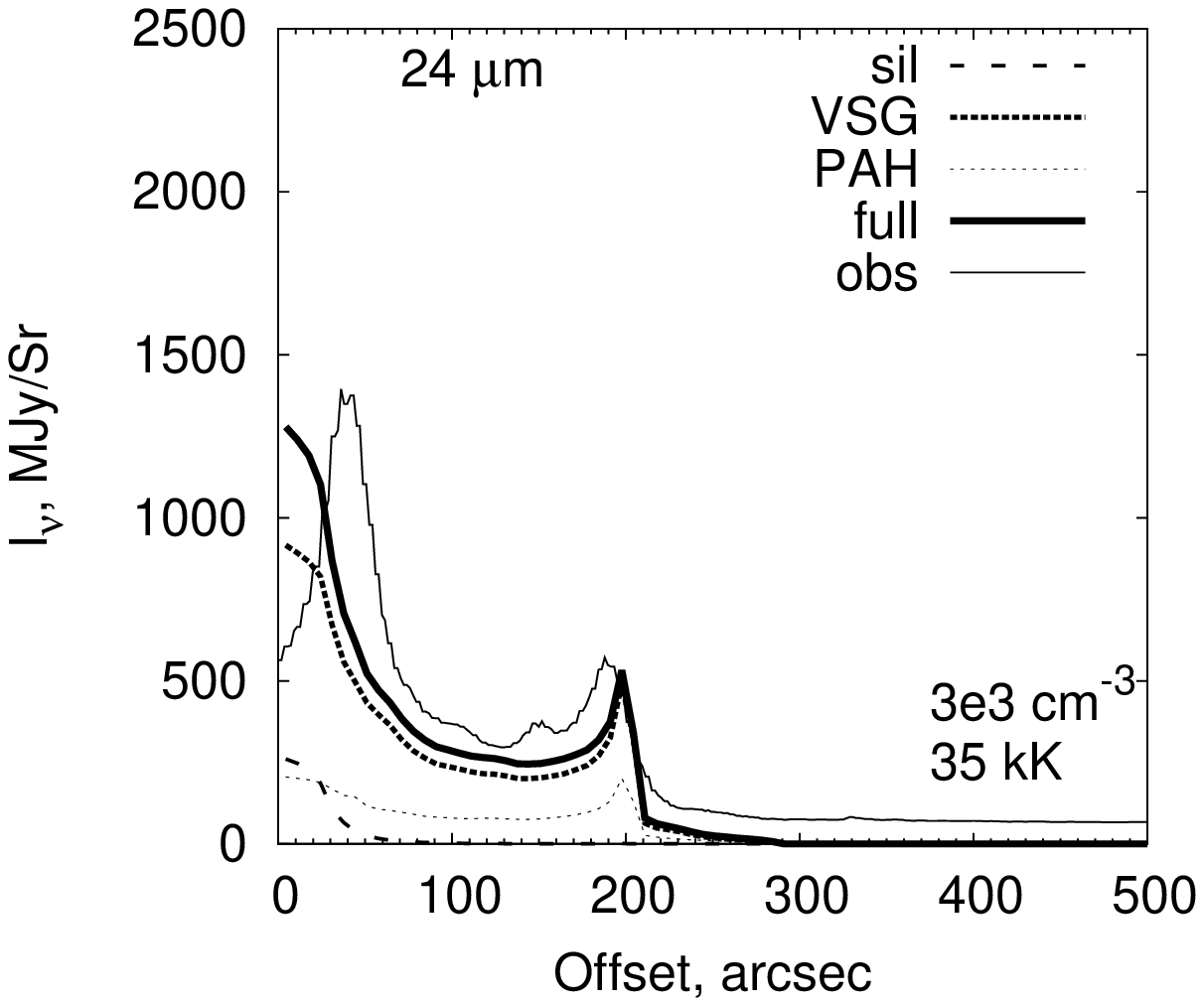}
\includegraphics[scale=0.42]{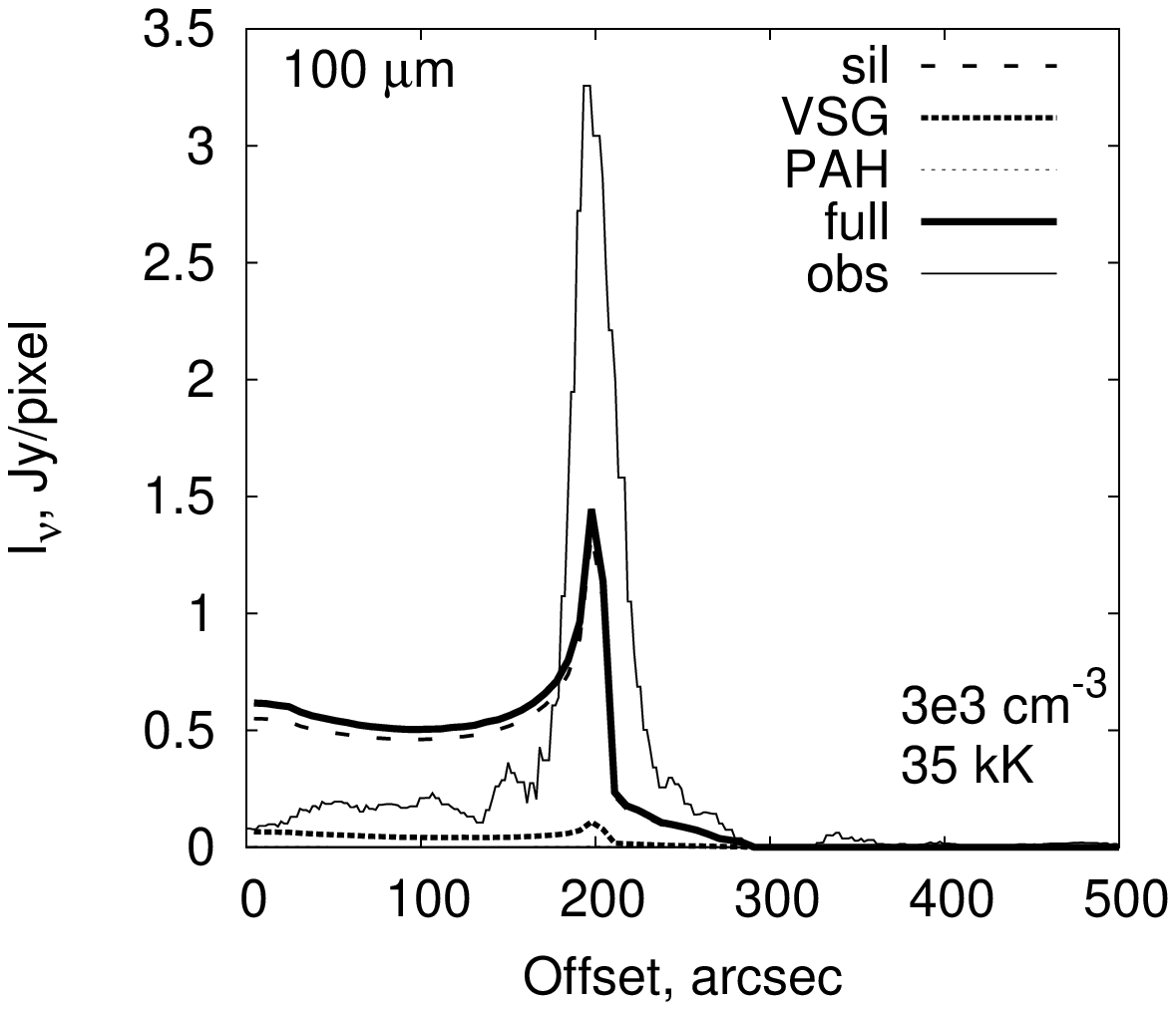}\\
\includegraphics[scale=0.42]{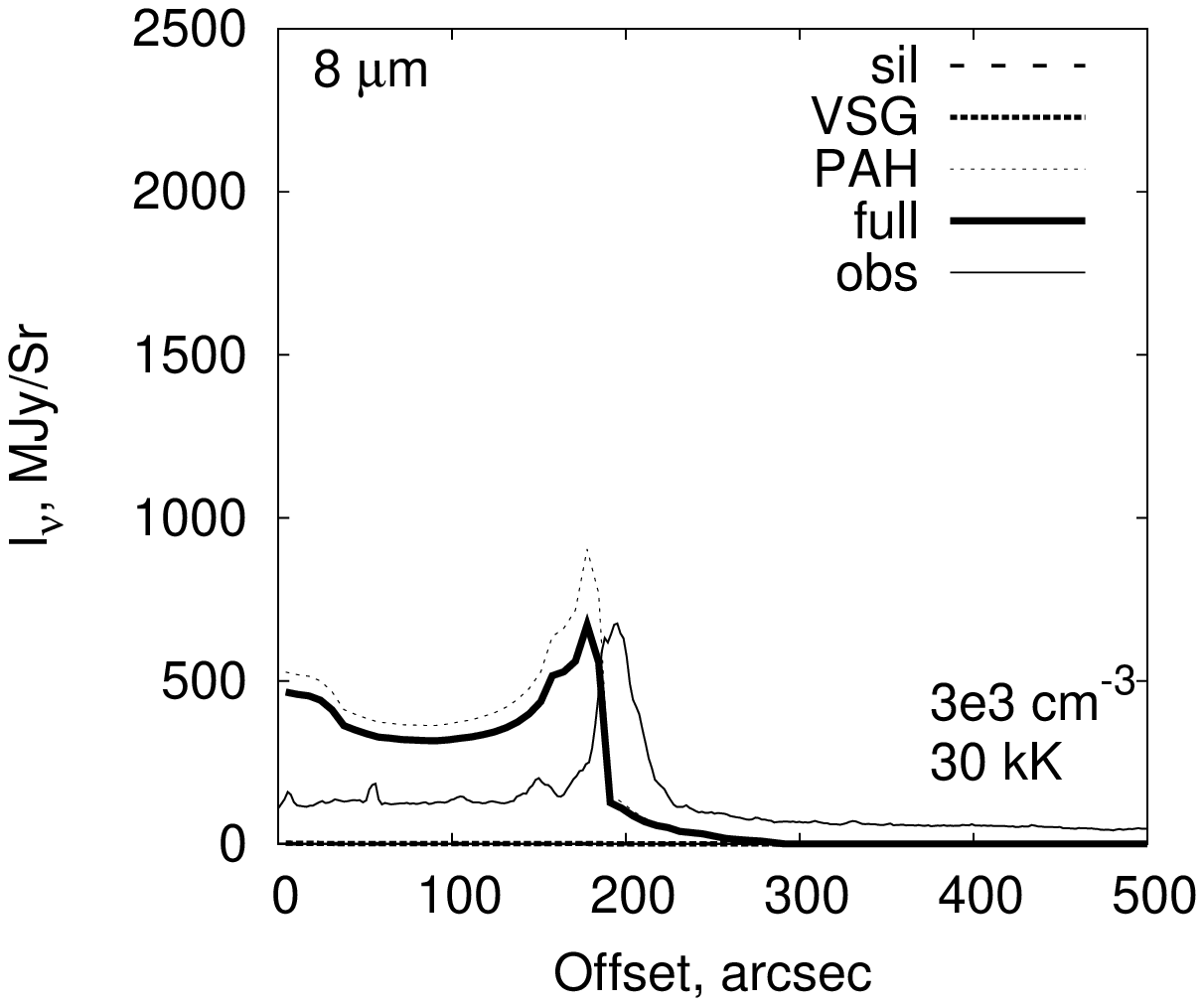}
\includegraphics[scale=0.42]{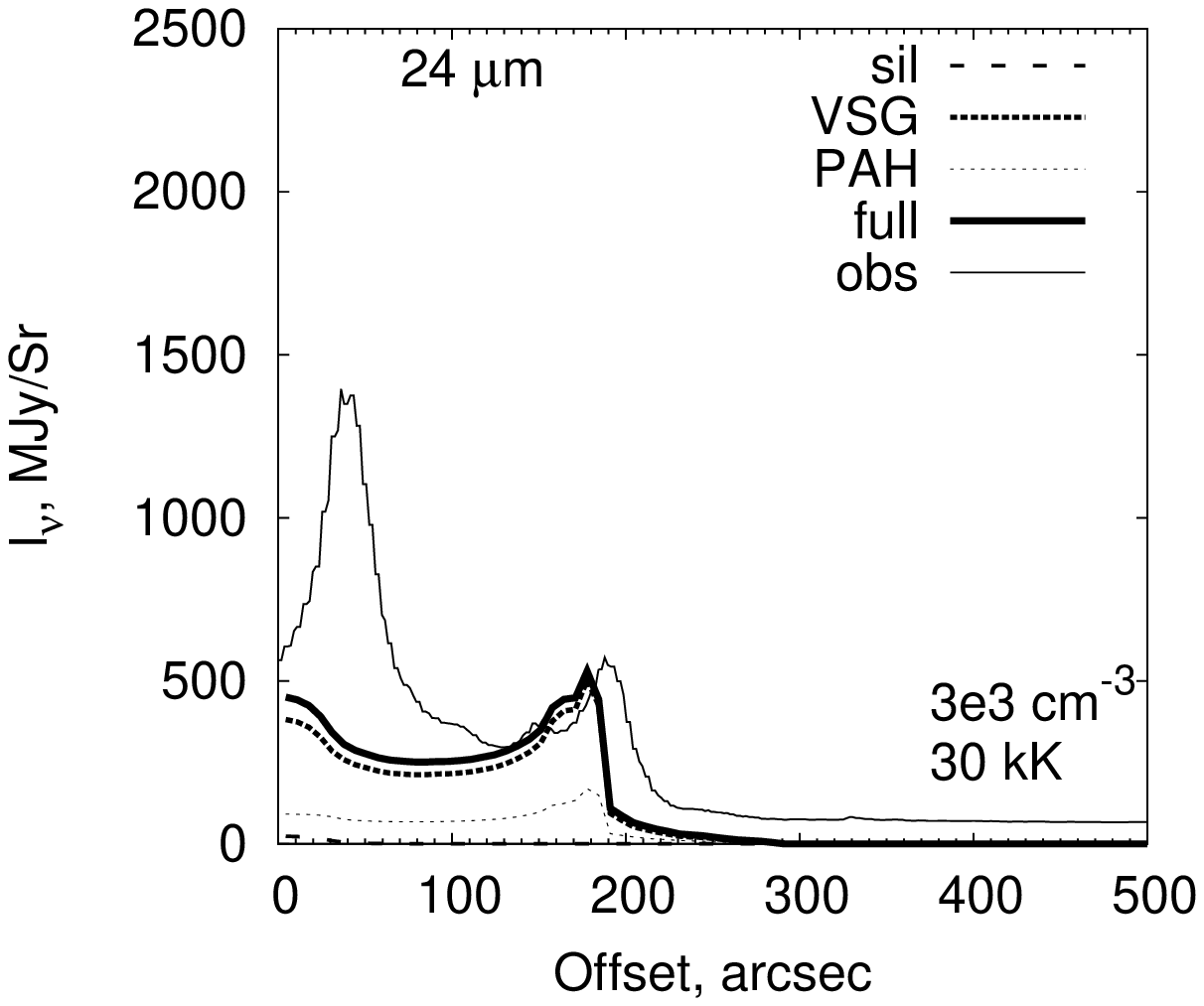}
\includegraphics[scale=0.42]{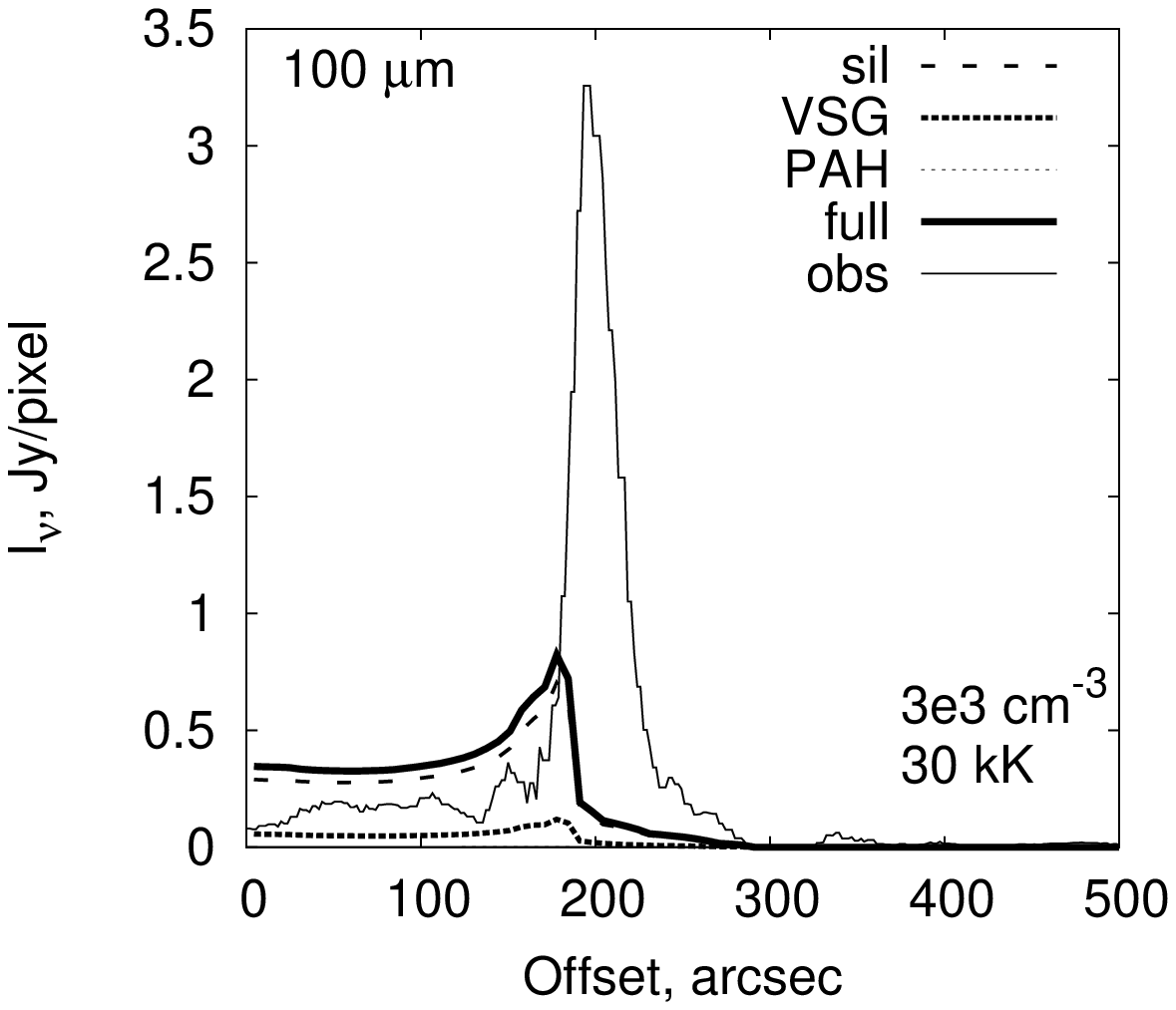}\\
\includegraphics[scale=0.42]{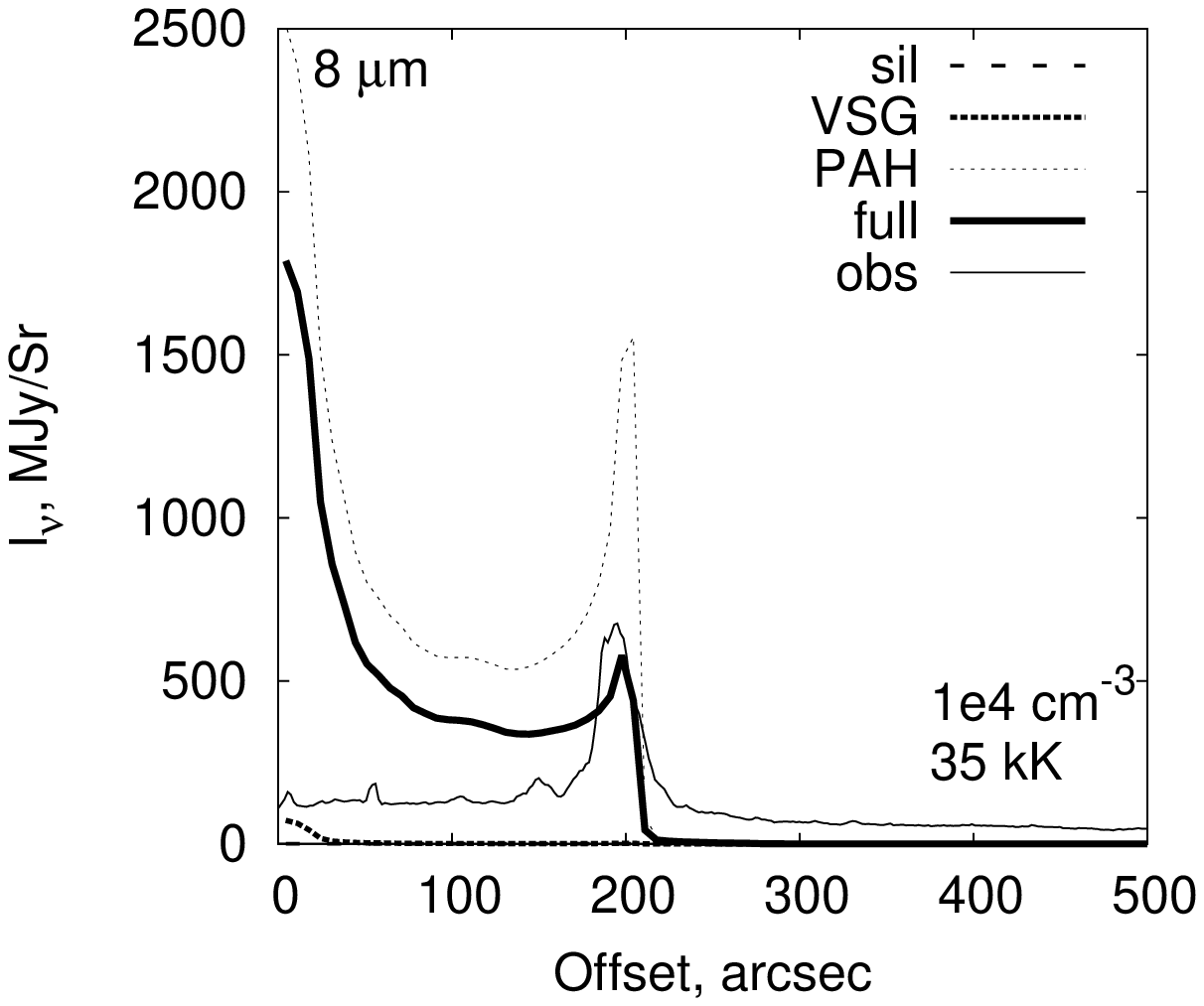}
\includegraphics[scale=0.42]{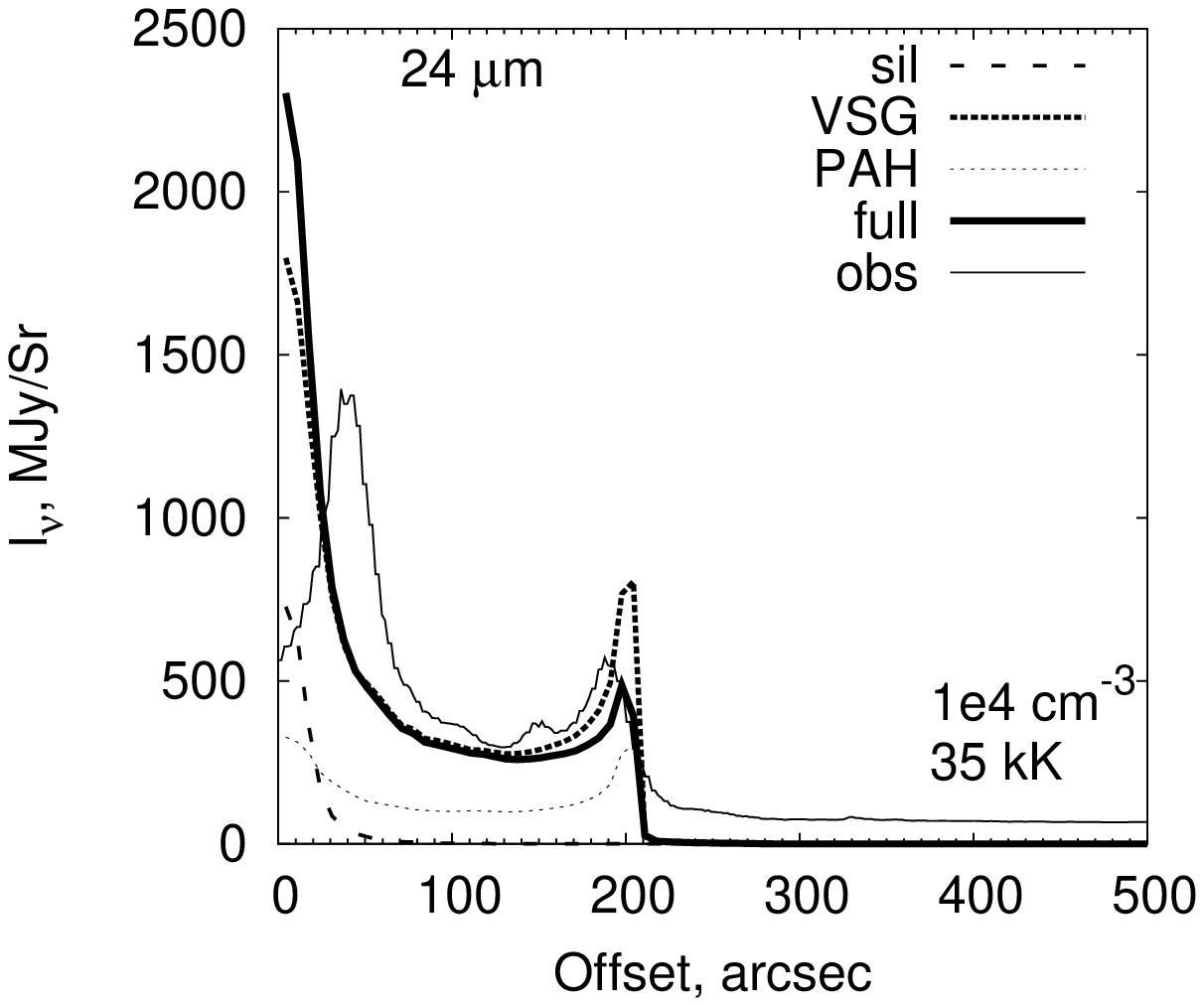}
\includegraphics[scale=0.42]{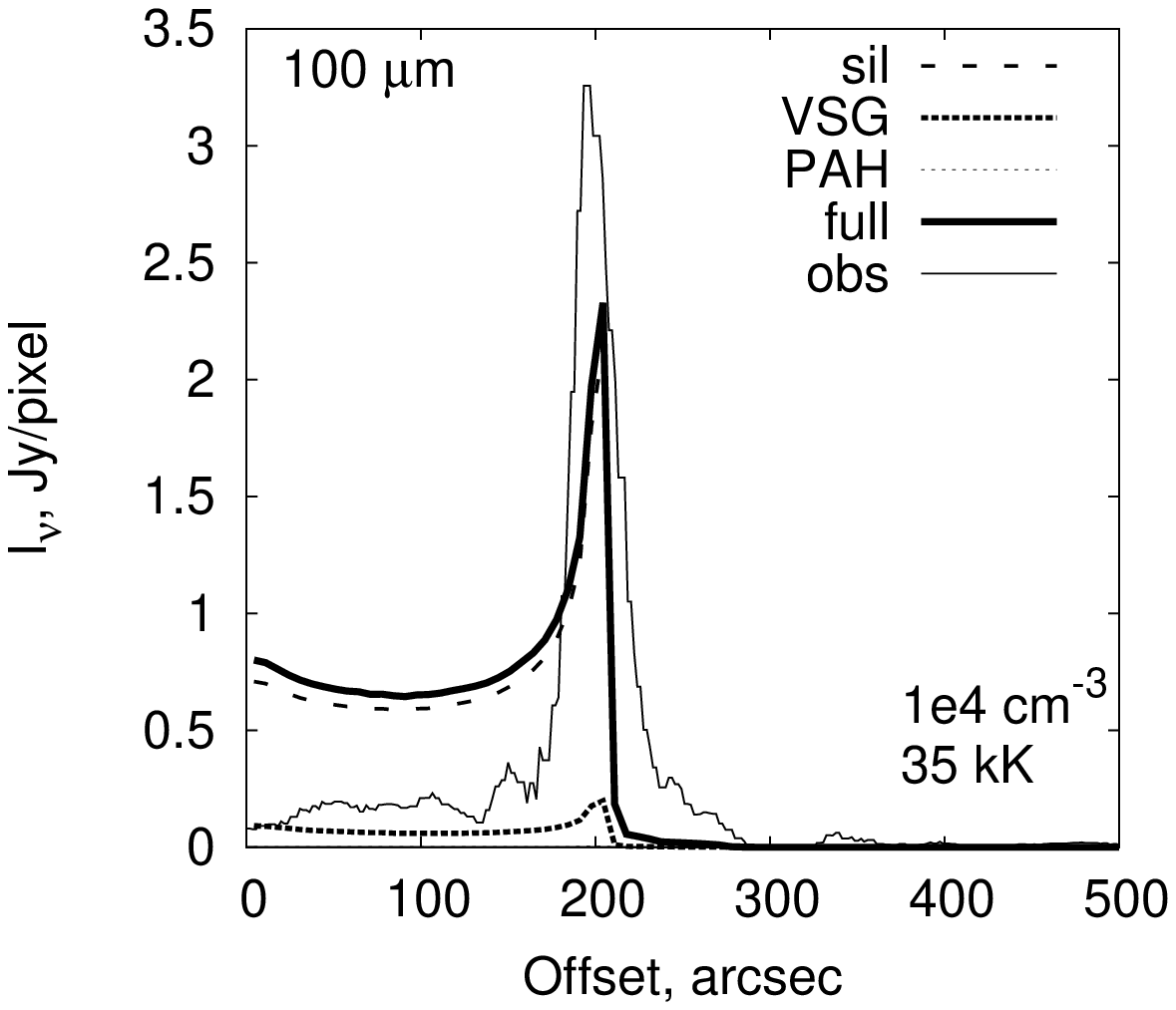}\\
\caption{Intensity distributions at 8\,$\mu$m (left column), 24\,$\mu$m (middle column), and 100\,$\mu$m (right column) for models with
various parameters. The thin solid curves (obs) show the observed profiles for RCW~120. The remaining notation is as in
Fig.~\ref{SED}.}
\label{mainfig}
\end{figure}

Figure~\ref{mainfig} shows the radial intensity distributions at three wavelengths, calculated using the models in Table\ref{nodestruction_models} and observed. The top row of panels presents results for the basic model, the middle row for the model with a reduced stellar temperature, and the bottom row for the model with an enhanced initial density.

The model with reduced stellar temperature (3e3-30) is in the worst agreement with the observations. Although the fraction of VSGs in this model has been artificially increased compared to the other two models in order to reproduce the intensity in the ring at 24\,$\mu$m, it predicts a substantially lower intensity toward the center of the object than is observed for RCW~120. The 100\,$\mu$m intensity in the direction toward the ring is also too low. When the stellar temperature is 30000~K, both the equilibrium heating of the large grains and the stochastic heating of the small grains are insufficient to provide the observed fluxes.

Increasing the stellar temperature by 5000 K fixes this problem for the fluxes at 24 and 100\,$\mu$m. In models 3e3-35 and 1e4-35, the small grains inside the HII region are heated sufficiently strongly to reproduce the observed increase in intensity at 24\,$\mu$m toward the center of the object. Note that our computations predict a peak in the emission toward the center, while the observed region of the 24\,$\mu$m maximum has the form of a narrow half-ring with radius 50 arcsec. It is possible that the distribution of dust in the central region of the HII region is influenced by both radiation and the stellar wind. The 100\,$\mu$m flux from the envelope is too low compared to the observations in both models, but this flux could be raised by refining the dust model used. We have not carried out such a refinement in this study, since it requires consideration of longer wavelength data.

A key property of all three models is an appreciable discrepancy with the observations at 8\,$\mu$m. With uniform distributions for all the dust components, the radial intensity profiles due to PAH emission and
to VSG emission are similar. In particular, in all three cases, we observe enhanced intensity toward the center of the object, with the center becoming brighter than the envelope at both 8\,$\mu$m and 24\,$\mu$m in models with a hotter star. While this picture is in agreement with observations at 24\,$\mu$m, there is no observed intensity enhancement in the central region of the HII region at 8\,$\mu$m. Consequently, a model with uniform distributions of all the dust components seems oversimplified, and it is necessary to take into account evolution of the dust. In the following section, we consider the influence of the destruction of PAH particles on the observed near-IR flux.

\subsection{RESULTS FOR MODELS WITH DESTRUCTION OF DUST}

We constructed an additional three models based on the basic model, in which PAH particles are destroyed by UV radiation from the star. These models differ in the characteristic time for the destruction of the PAH particles in a standard interstellar radiation field, $\tau_{\rm PAH}$. We considered the three values $\tau_{\rm PAH} =$~3, 30, and 300 million years. (We have not considered here models with the destruction of VSGs; i.e., in all cases we assume $\tau_{\rm VSG}=\infty$.) The notation for these
models is given in Table~\ref{pahdestruction_models}.

\begin{table}[h!]
\caption{Models with destruction of PAH particles}
\bigskip
\begin{tabular}{l|c}
\hline
Model & $\tau_{\rm PAH}$, million years \\
\hline
3e3-35-3 & 3 \\
3e3-35-30 & 30 \\
3e3-35-300 & 300 \\
\hline
\end{tabular}
\label{pahdestruction_models}
\end{table}

\begin{figure}[t!]
\includegraphics[scale=1.0]{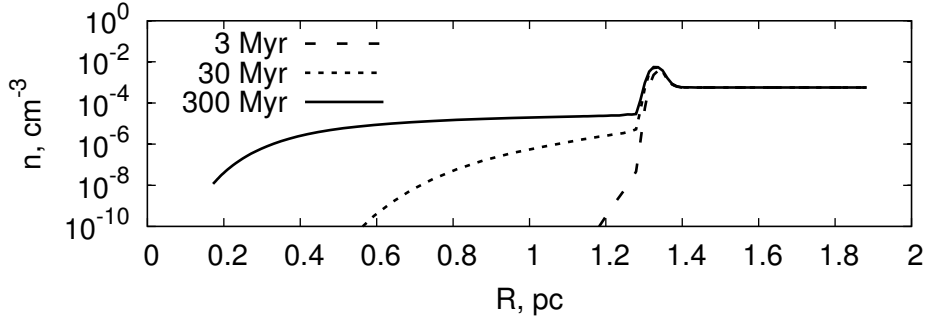}
\caption
{Radial distributions of the PAH number density for various characteristic destruction times $\tau_{\rm PAH}$. Results for the basic model, corresponding to an age of the HII region of $t = 170 000$~yrs, are shown.}
\label{dusdes}
\end{figure}

The radial distributions of the PAH number density in the basic model for the HII region are shown in Fig.~\ref{dusdes}. The characteristic time for the PAH destruction in the radiation field of the central star in model 3e3-35-300, $\tau_{\rm PAH}/G\sim1$ million years, exceeds the age of the HII region corresponding to the observed size of RCW120 (170 000 years); therefore, the PAH density differs from the results of the basic model without dust destruction only in the very center of the HII region. Other aspects of the structure of the HII region remain virtually unchanged in the model with PAH destruction, and we accordingly analyse further only the theoretical radial IR intensity profiles.

\begin{figure}[t!]
\includegraphics[scale=1.0]{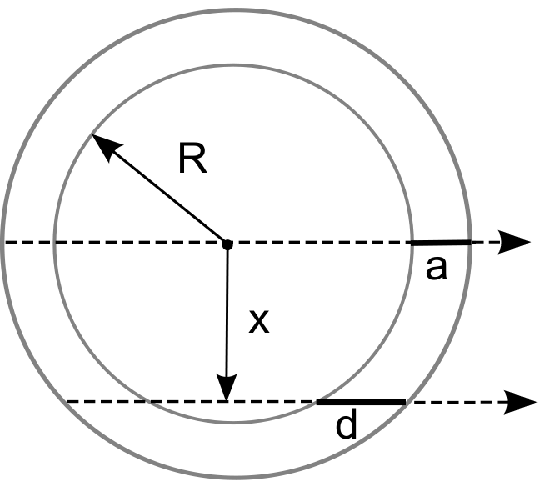} 
\includegraphics[scale=0.4]{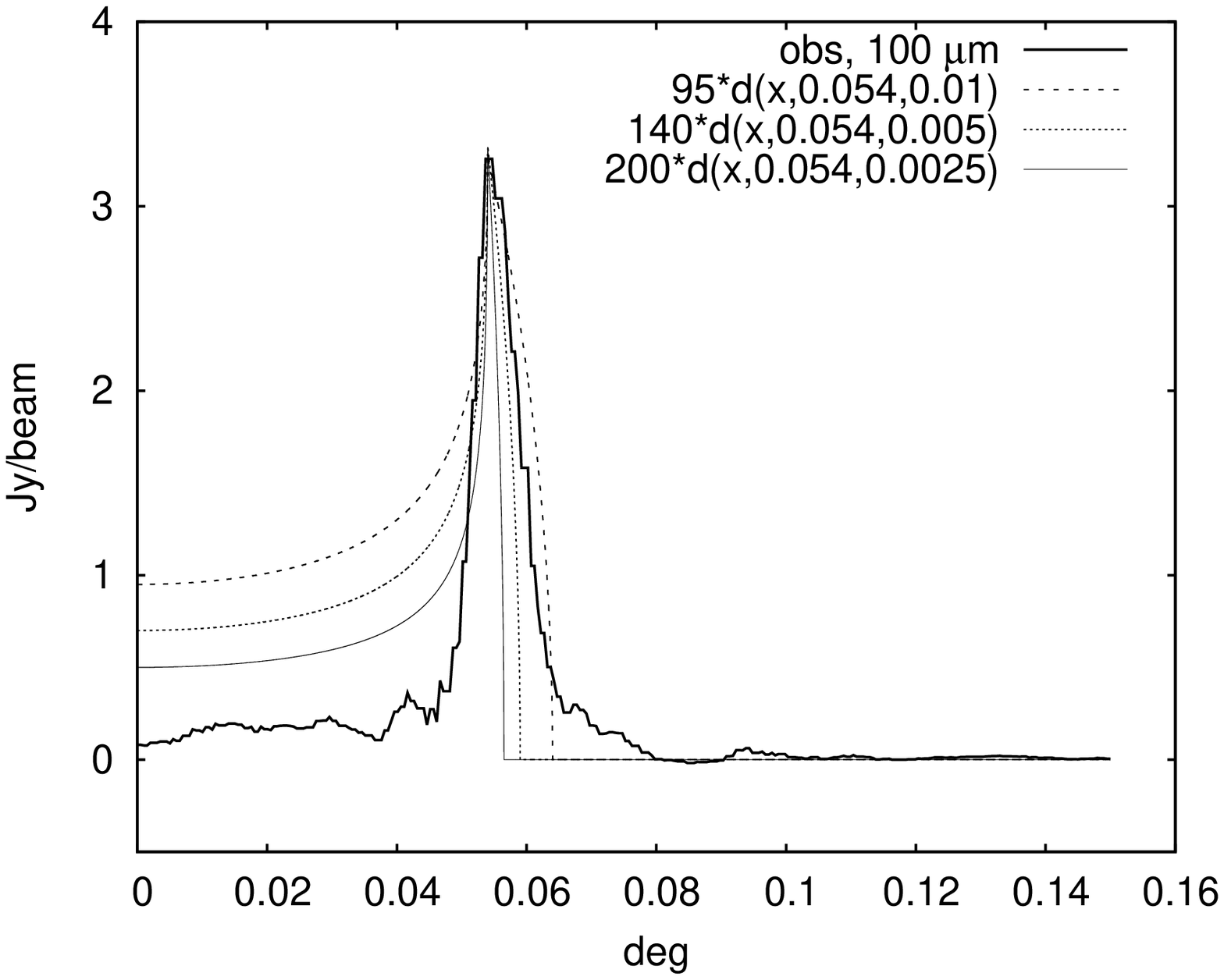} 
\caption{Left: schematic for the computation of the intensity distribution for a spherical layer lying between radii $R$ and $R + a$. Right: normalized profiles of the function $d(x)$ for various values of $R$ and $a$. The bold solid curve (obs) shows the observed intensity distribution at 100\,$\mu$m for RCW~120.}
\label{scheme}
\end{figure}

Let us first present some simple geometrical reasoning about the expected form of these distributions. Let us consider a spherical layer lying between radii $R$ and $R + a$ (Fig.~\ref{scheme}, left), and find the intensity distribution in the plane of the sky, assuming that emission was generated only in this layer. In our formulation, this layer corresponds to the dense shell of the HII region, inside of which there are no sources of emission (PAH particles).

In an optically thin approximation, the intensity at an offset corresponding to an impact parameter of $x$ is proportional to the length of the intersection of the line of sight and the layer, i.e., the length $d$. This length (within the near hemisphere) can be calculated using the formula
\begin{equation}
d(x,R,a)=\left\{
\begin{array}{ll}
\sqrt{(R+a)^{2}-x^2}-\sqrt{R^{2}-x^2}, & x<R \\
\sqrt{(R+a)^{2}-x^2}, & R<x<R+a \\
0, & x>R+a
\end{array}
\right.
\end{equation}
The right panel in Fig.\ref{scheme} presents normalized profiles $d(x,R, a)$ for various values of $R$ and $a$, together with the 100\,$\mu$m intensity distribution for RCW~120. This figure shows that an optically thin, spherically symmetrical layer should display appreciable emission not only in a ring, but also toward the center. Further, we compare this general conclusion with the results of
our computations and observations.

\begin{figure}[t!]
\includegraphics[scale=0.42]{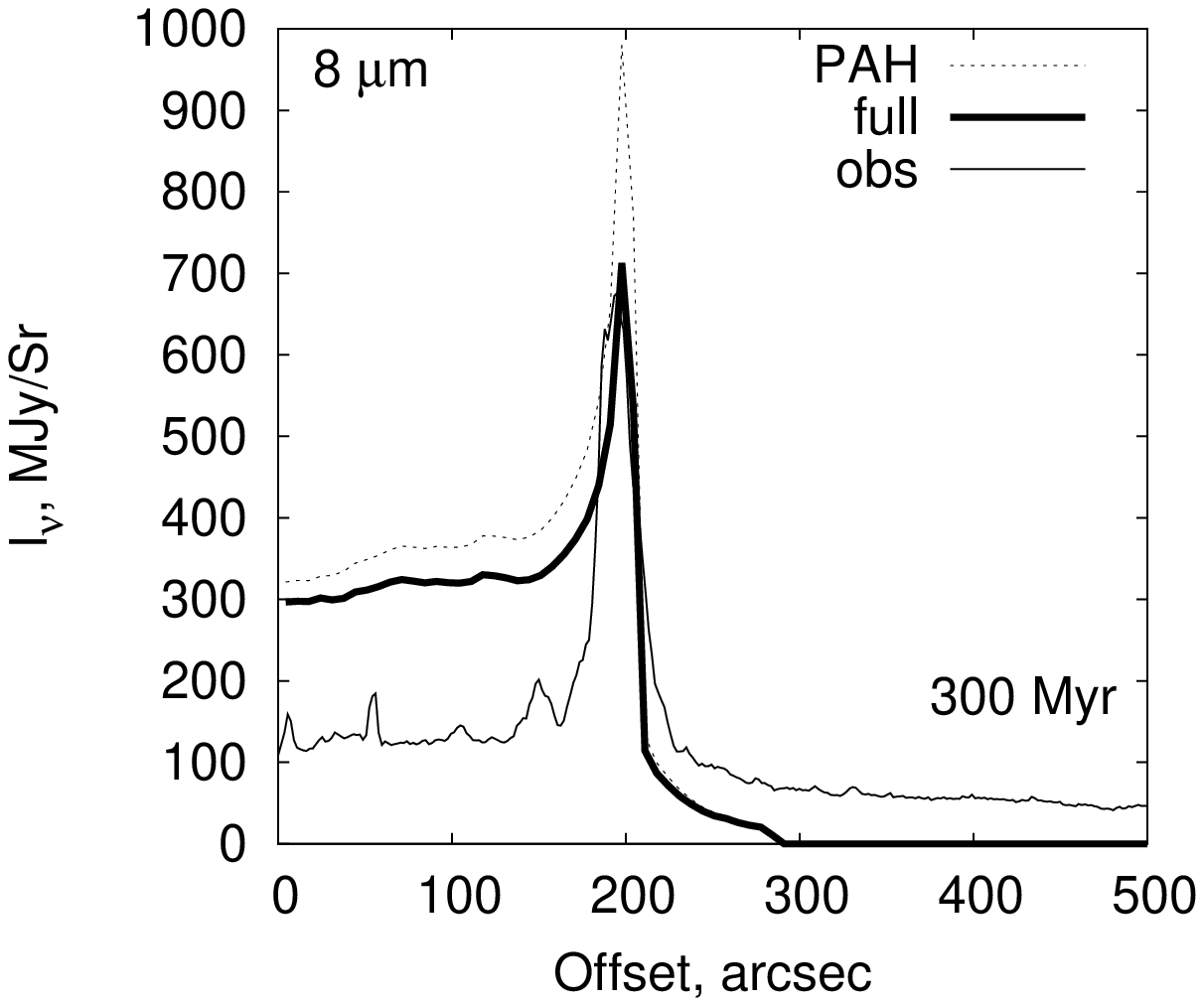}
\includegraphics[scale=0.42]{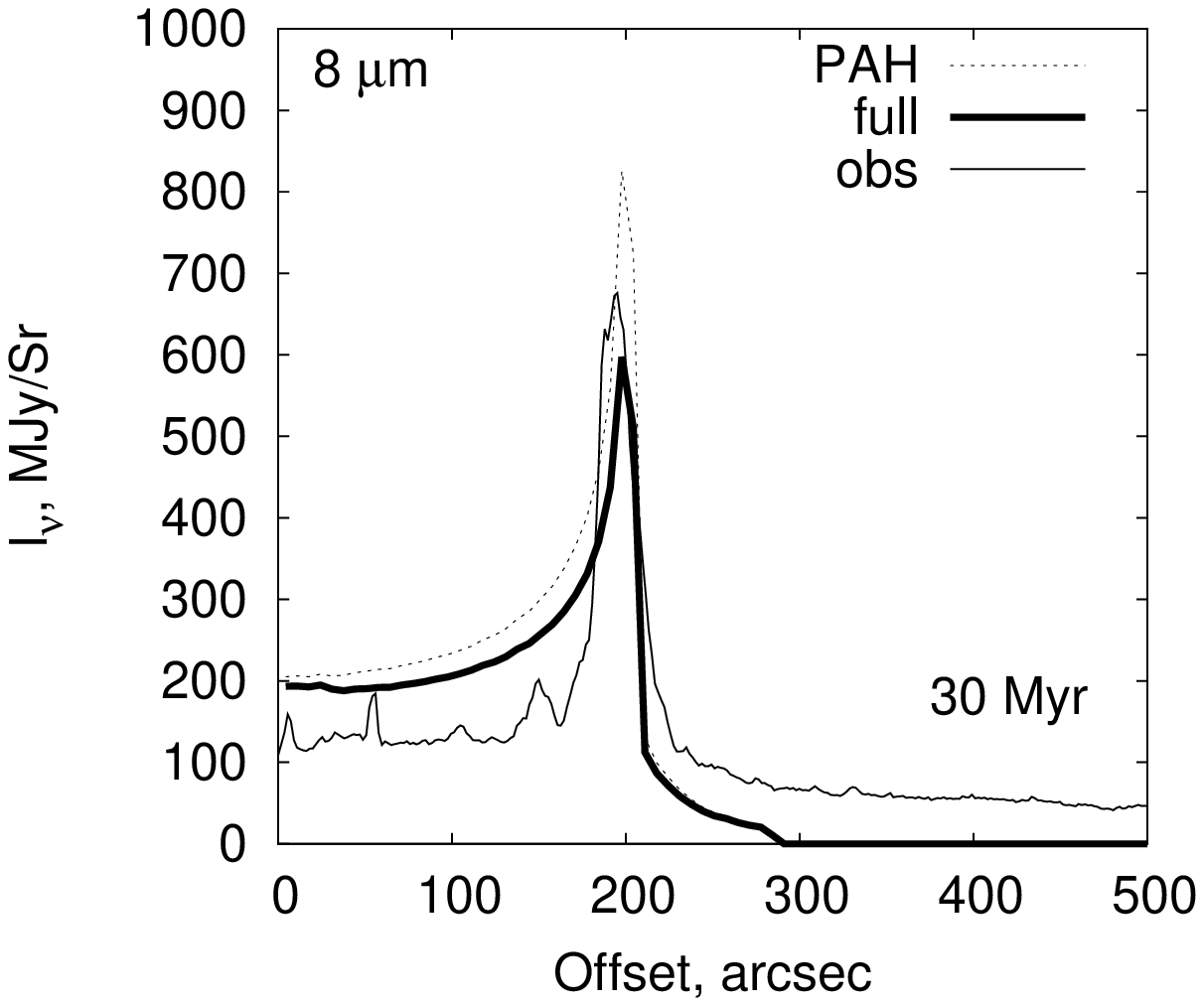}
\includegraphics[scale=0.42]{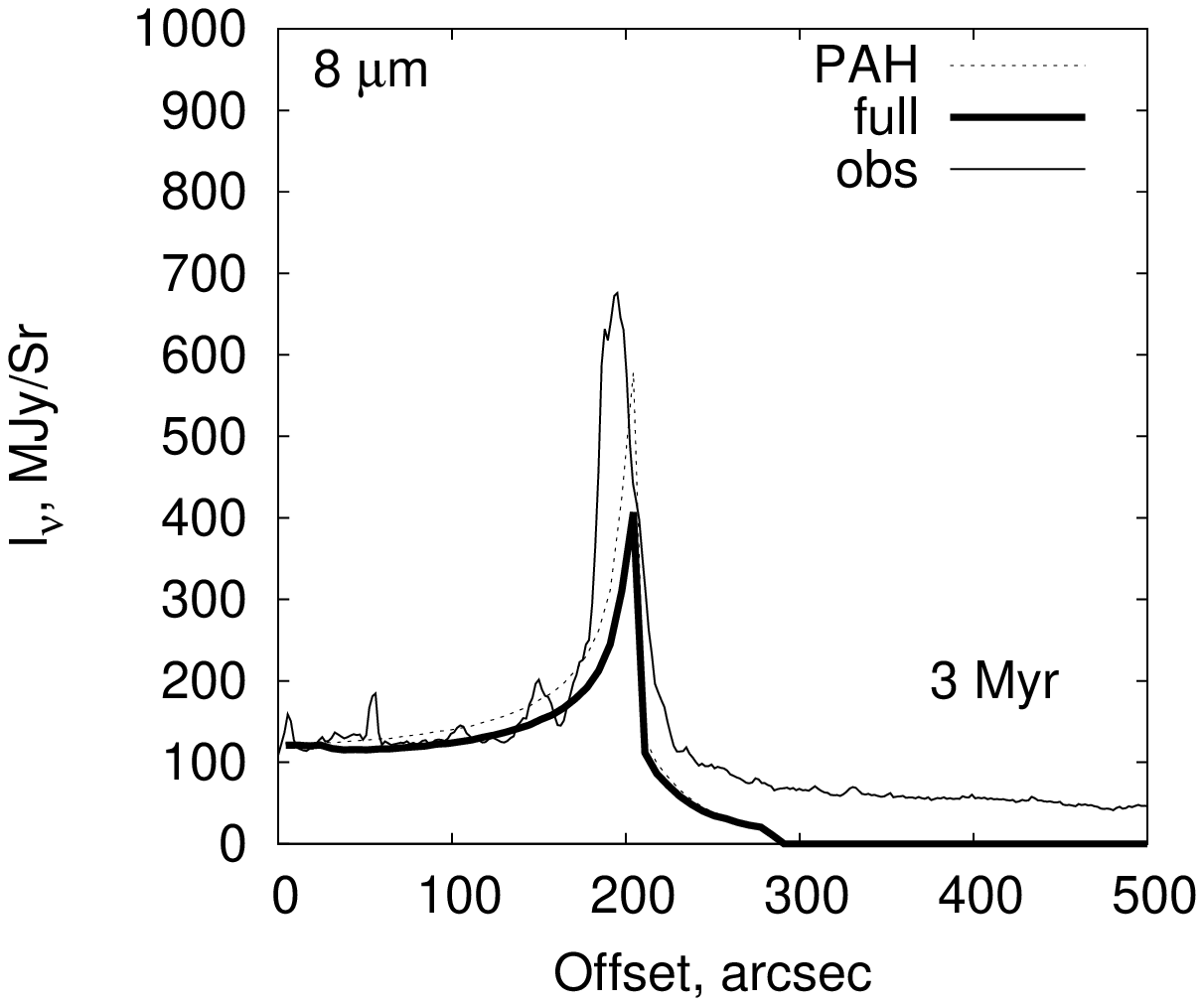}
\caption{8\,$\mu$m intensity distribution for models with various characteristic PAH-destruction times. The bold, solid curves (total) show the model profiles and the thin, solid curves (obs) the observed profile for RCW~120. The dotted curves show the contribution of the PAH emission. The closeness of the dotted curve and thin, solid curve indicates that essentially all the emission at this wavelength is generated by PAH particles.}
\label{mainfig1}
\end{figure}

Figure~\ref{mainfig1} shows radial profiles of the 8-$\mu$m intensity for the models listed in Table~\ref{pahdestruction_models}. It is obvious that even slow destruction of the PAHs (model 3e3-35-300), leading to a decrease of the PAH concentration only at the very center of the HII region, solves the problem of the central peak: compare the upper, left panel of Fig.~\ref{mainfig} and the left panel of Fig.~\ref{mainfig1}. However, the overall level of the emission in the center of the HII region is too high in this model. Model 3e3-35-3, with the most rapid destruction of PAH particles, can reproduce the observed flux at the center of the object, but the emission in the ring becomes too weak compared to the observations. Overall, the best
agreement with the observations is provided by the model with $\tau_{\rm PAH}=30$ million years.

The form of the radial intensity profile in this model fully corresponds to our theoretical expectations: most of the emission aries in a narrow shell whose thickness does not 0.1~pc, but this emission is sufficient to provide an appreciable enhancement above the background, both toward the shell and toward the center of the object, where the envelope thickness is minimum.

\section{DISCUSSION AND CONCLUSIONS}

We have considered the IR emission from dust in a region of ionized hydrogen. The observations show that the emission from such objects is spatially separated at 8 and 24\,$\mu$m. Our computations indicate that this spatial separation can be explained only if some of the dust components, in particular, the PAHs, are absent in the region of ionized hydrogen, and are concentrated in a narrow shell surrounding this region. It was suggested earlier that the absence of dust in an HII region could be associated with the action of radiation pressure or the stellar wind. Considering the HII region RCW~120, we have concluded that the observed radial emission profile at 8\,$\mu$m can be reproduced if the PAH particles are destroyed by UV radiation, with a characteristic destruction time in a standard interstellar radiation field of the order of 30 million years.

It is important to note that the observed 24-$\mu$m profile can be reproduced if VSGs are present in the HII region in the same proportion as in the unionised matter; i.e., the mechanism responsible for eliminating the PAHs from the HII region must not affect the larger particles. The action of the stellar wind on the VSGs should be appreciable only in the immediate vicinity of the star, where our model predicts a central peak in the 24-$\mu$m emission, while the radiating region has the form of a compact ring~\cite{evch}.

The situation with regard to the IR emission at 100\,$\mu$m is somewhat more complex. All the models we have considered predict appreciable emission at the center of the object, while this emission is essentially absent in the observations. The \textit{Herschel} data at this wavelength indicate an intensity toward the center that is virtually indistinguishable from the background. A similar picture is
observed at wavelengths longer than those accessible to \textit{Herschel}. Geometrical reasoning shows that this problem cannot be solved by supposing an absence of large grains in the region of ionized hydrogen. Even a narrow shell should lead to enhanced emission toward the center of the object.

Two explanations are possible here. First, it may be that the HII region of RCW~120 is not spherical, but instead cylindrical, and that we are observing the object along the axis of this cylinder. Second, it is possible that the reduced emission at the center of RCW~120 is associated with peculiarities of the reduction algorithm applied for the data in the \textit{Herschel} science archive~\cite{rcw120hersch}. We plan to test these ideas by considering several other similar objects and carrying out our own reduction of the data at 100\,$\mu$m and longer wavelengths.

This study is a first step in our numerical investigation of the destruction of dust in regions of ionized hydrogen. The most important directions for the further development of our model include: studies of aspects of the chemical evolution of an HII region associated with the destruction of PAHs; consideration of not only the current state, but also evolutionary variations, especially possible observational manifestations of early stages in the evolution of an HII region; and investigation of the character of the destruction of dust particles in HII regions around less massive stars. We intend to pursue these directions in our subsequent studies.

\begin{acknowledgments}
This work was supported by the Russian Foundation for Basic Research (projects 10-02-00612 and 12-02-31248), the Program of State Support for Leading Scientific Schools of the Russian Federation (grant NSh-3602.2012.2), and the Basic Research Program of the Presidium of the Russian Academy of Sciences P-21, "Non-stationary Phenomena in
Objects of the Universe."

\textit{Herschel} is an ESA space observatory with science instruments provided by European-led Principal Investigator consortia and with important participation from NASA. This work is based in part on archival data obtained with the \textit{Spitzer} Space Telescope. This research has made use of the NASA/IPAC Infrared Science Archive. The telescope and archive are operated by the Jet Propulsion Laboratory, California Institute of Technology under a contract with NASA.
\end{acknowledgments}

\end{document}